\documentclass[reprint,aip,apl]{revtex4-1}
\usepackage{graphicx,charter,hyperref,braket}

\begin{document}

\raggedbottom

\title[Weyl nodes in Mn$_3$ZnC]{Weyl nodes and magnetostructural instability in 
antiperovskite Mn$_3$ZnC}

\author{S. M. L. Teicher}
\affiliation{Materials Department and Materials Research Laboratory, University of California, Santa Barbara 93106, USA}
\email{steicher@ucsb.edu}
\author{I. K. Svenningsson}
\affiliation{Materials Research Laboratory, University of California, Santa Barbara 93106, USA}
\affiliation{Department of Applied Physics, Chalmers University of Technology 412 96, Sweden}
\author{L. M. Schoop}
\affiliation{Department of Chemistry, Princeton University, Princeton 08540, USA}
\author{R. Seshadri}
\affiliation{Materials Department and Materials Research Laboratory, University of California, Santa Barbara 93106, USA}

\date{\today}

\begin{abstract}
The room temperature ferromagnetic phase of the cubic antiperovskite Mn$_3$ZnC is suggested from first-principles
calculation to be a nodal line Weyl semimetal. Features in the electronic 
structure that are the hallmark of a nodal line Weyl state---a large density of linear band crossings near the Fermi level---can also be interpreted as 
signatures of a structural and/or magnetic instability. Indeed, it is known that 
Mn$_3$ZnC undergoes transitions upon cooling from a paramagnetic to a cubic ferromagnetic state under ambient conditions
and then further into a non-collinear ferrimagnetic tetragonal phase at a temperature
between 250\,K and 200\,K. The existence of Weyl nodes and their destruction via
structural and magnetic ordering is likely to be relevant to a range of magnetostructurally 
coupled materials.
\end{abstract}


\maketitle

\section{Introduction}

\begin{figure*}
\includegraphics[width=0.9\textwidth]{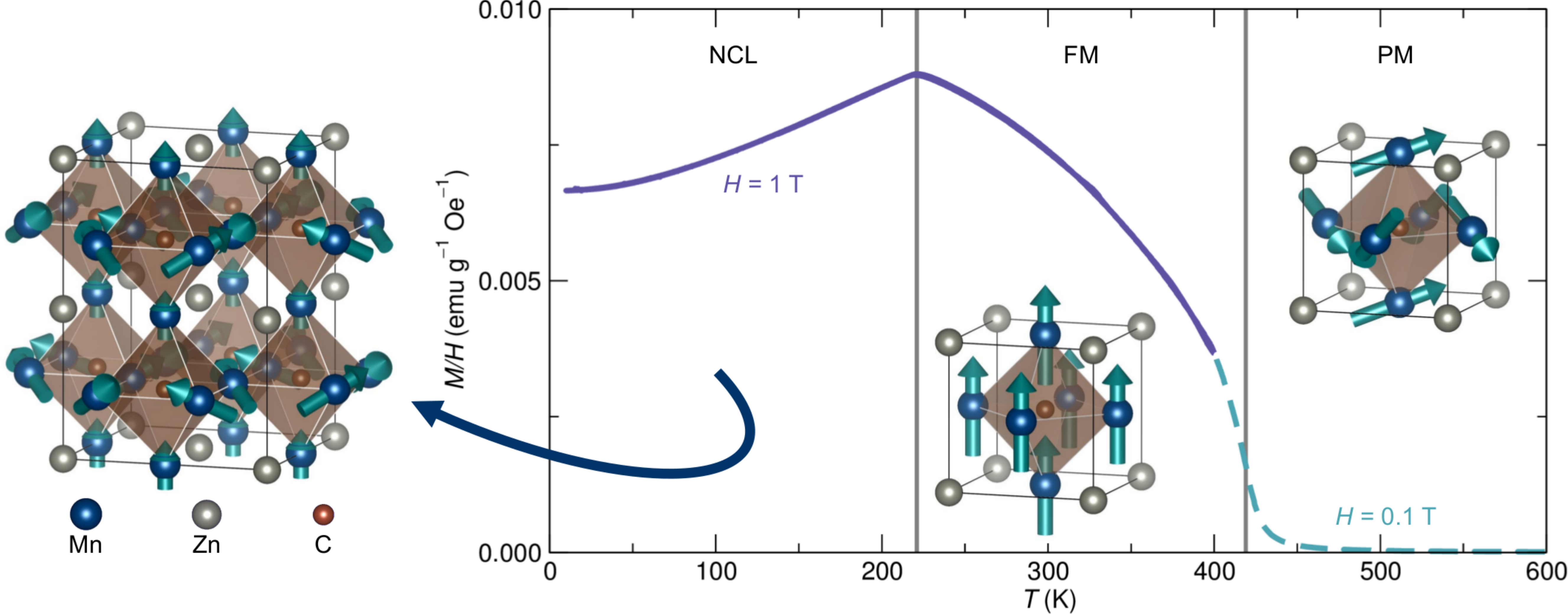}
\caption{\label{fig:Figure1} Reported crystal and magnetic structures of Mn$_3$ZnC. Mn$_3$ZnC undergoes
transitions from paramagnetic (PM) to ferromagnetic (FM) states in the cubic antiperovskite structure with 
a $T_C\approx$420\,K. A lower temperature antiferromagnetic transition with $T_N\approx$219\,K is 
associated with noncollinear ferrimagnetic ordering in a tetragonal structure.}
\end{figure*}

Antiperovskite carbides are a family of materials with the cubic perovskite structure and formula $X_3B$C where $X$ 
is the least electronegative element in the formula and C is carbon (see Fig.\,\ref{fig:Figure1}). Related carbides have a 
long history in metallurgical research, associated with the larger family of so-called $MAX$ phases.\cite{Barsoum2000} 
Several members of the family such as Co$_3$AlC\cite{Kimura2001} have been explored for applications in structural alloys and Mn$_3$SnC\cite{Wen2009} and carbon-doped Mn$_3$ZnN\cite{Hamada2011} exhibit negative thermal expansion effects. 
Some antiperovskite carbides are also of interest for their functionality. Mn$_3$GaC is known to display giant 
magnetoresistance\cite{Kamishima2000} and Ni$_3$MgC is an 8\,K superconductor, the latter rationalized by first principles calculations 
as being associated with a flat band with large density of states near the Fermi level.\cite{He2001,Mollah2004}

More recently, the larger class of antiperovskites has been explored due to the prediction of topological electronic 
states. The Ca$_3$PbO and Ca$_3$BiN families include Dirac semimetals\cite{Kariyado2011,Kariyado2012}---materials with a graphene-like linear-band crossing, or Dirac cone, near the Fermi level\cite{CastroNeto2009,Yang2014}---as well as topological insulators\cite{Goh2018} and topological crystalline insulators\cite{Hsieh2014} with a bulk band gap and metallic surface states. Magnetization and resistivity data in two recent studies on Sr$_3$PbO and Sr$_3$SnO provide preliminary evidence for Dirac transport\cite{Suetsugu2018} and low-temperature superconductivity,\cite{Oudah2016} respectively. The Cu$_3$PdN family is predicted to include nodal line semimetals, in which Dirac crossings persist over an extended region, a line, rather than a single point in the Brillouin zone.\cite{Yu2015} Nodal line semimetals can host a unique class of drum-head topological surface states with $k$-vectors connecting all the Dirac nodes on the nodal ring in the Brillouin zone.\cite{Schnyder2016} However, as is common in predicted nodal line compounds, the relatively large 
spin-orbit coupling due to the heavy Pd atom in Cu$_3$PdN partially gaps the nodal line, preventing the realization of 
a true nodal line semimetal.\cite{Yu2015}

The topic of this letter is Mn$_3$ZnC, a material that has been explored since the 1950s due to its interesting magnetic 
transitions. In a series of studies, Butters and Myers,\cite{Butters1955a} Brockhouse and Myers,\cite{Brockhouse1957} and Swanson and Friedberg\cite{Swanson1961} provided early characterization of these transitions, which include a paramagnetic to ferromagnetic transition with reported Curie temperatures 350\,K$<T_C<$500\,K\cite{Butters1955a,Fruchart1978,Kaneko1987a} and antiferromagnetic ordering with reported N{\'e}el temperatures 215\,K$<T_N<$233\,K.\cite{Butters1955a,Fruchart1978,Kaneko1987a} In the 1970s, Fruchart and colleagues solved the magnetic structures via neutron diffraction, determining the [001] ferromagnetic room temperature structure and more complex non-collinear low temperature structure shown in Fig \ref{fig:Figure1}.\cite{Fruchart1973,Fruchart1978} The magnetic moments, localized on the Mn site, are approximately 1.3\,$\mu_B$ in the ferromagnetic state and 2.7$\mu_B$ and 
1.6\,$\mu_B$ in the noncollinear, antiferromagnetically-coupled and collinear, ferromagnetically-aligned layers of the low 
temperature structure, respectively. The low magnetic moments suggest that the magnetism is fairly itinerant in this system relative to 
most Mn magnetic materials. In the 1980s, the transition shifts in Mn$_3$ZnC were studied under large 
applied magnetic fields and pressures.\cite{Kaneko1987a,Kaneko1987b}

Theoretical attempts to explain the magnetic transitions in Mn$_3$ZnC have focused on the presence of a large 
Fermi level flat band. In 1975, Jardin and Labb{\'e} proposed that the main bonding interactions in this system 
originate from Mn $d_{xz},d_{yz}$ and C $p$ orbitals and generated a tight-bonding model with sharp peaks in 
the density of states.\cite{Jardin1975} Should the Fermi level lie at one of these peaks, the ferromagnetic 
to paramagnetic transition could be explained by Stoner exchange and the low temperature structural distortion 
by a Jahn-Teller-like energetic benefit accrued through further reducing the density of states near the Fermi level. 
Non-spin polarized density of states calculations reported later indeed found a large spike in the 
density of states just above the calculated Fermi level.\cite{Motizuki1988} This work was extended with density of states calculations on both the spin-polarized, ferromagnetic and non-collinear, ferrimagnetic structures, demonstrating that spin polarization opens up a pseudo gap at the Fermi-level in the ferromagnetic phase, consistent with a Stoner exchange mechanism, and that a wider gap opens just below $E_F$ in the ferrimagnetic state.\cite{Antonov2007}

Here, we use first principles calculations to show that the cubic ferromagnetic phase of Mn$_3$ZnC that is stable at room temperature is a nodal line Weyl semimetal. Although Weyl nodes are common in ferromagnetic metals, including bcc Fe,\cite{GosalbezMartinez2015} to-date studies on Weyl nodal lines are more of a rarity: the first experimental work towards verification of a Weyl nodal line has recently been completed for the Heusler compound MnCo$_2$Ga.\cite{Belopolski2019} In Mn$_3$ZnC, we present a computational bonding analysis that provides qualitative explanations for the magnetic transitions that 
create and destroy this delicate Weyl ferromagnet phase. Motivated by interest in the Weyl nodes and their potential role in 
magnetostructural instabilities, we experimentally revisit the transitions in this material via magnetoentropic mapping,  
which is a sensitive probe for low energy and metamagnetic reordering. 

\section{Methods}

\subsection{Computational}

Density functional theory simulations were completed in VASP \cite{Kresse1994,Kresse1996a,Kresse1996b} and WIEN2k\cite{Blaha2001} using the Perdew, Burke, Ernzerhof functional \cite{Perdew1996} with projector-augmented waves,\cite{Blochl1994a,Kresse1999} and linear augmented plane waves $+$ local orbitals\cite{Schwarz2002}, respectively. PAW potentials for VASP were selected based on the version 5.2 recommendations. All calculations described in the text were performed in VASP other than the band irrep assignments of Fig.\,\ref{fig:Figure3} (c), which were completed using the \textsc{irrep} subprogram of WIEN2k. A $17\times17\times17$ $\Gamma$-centered $k$-mesh was employed for electronic structure simulations of the ferromagnetic phase. These calculations are well-converged for much lower density $k$-meshes, but high $k$-mesh density is desirable for accurate Wannier function fitting; an $8\times8\times8$ mesh was used for relaxation steps. Calculations for the expanded low temperature tetragonal cell were performed using a $5\times5\times4$ $\Gamma$-centered $k$-mesh for relaxation and self-consistent steps and a $7\times7\times5$ mesh for the density of states. The plane wave energy cutoff for VASP and the plane-wave expansion parameter, RKMAX, for WIEN2k were set to values better than 500\,eV and 8.0, respectively. Tetrahedral smearing with Bl{\"o}chl corrections\cite{Blochl1994b} was used for relaxations and self-consistent calculations. Structures were relaxed in VASP via the conjugate gradient descent algorithm with an energy convergence cutoff of $10^{-4}$\,eV. Subsequent self-consistent static calculations and non self-consistent electronic structure calculations were performed using VASP and WIEN2k with energy convergence better than $10^{-5}$\,eV. The surface states of Mn$_3$ZnC were determined by projecting our VASP calculations onto maximally localized Wannier functions using Wannier90,\cite{Mostofi2014} starting from initial projectors corresponding to valence orbitals (Mn d; Zn s, p, d; C p; with a frozen fitting window $E_F \pm 2$\,eV), constructing a tight-binding model from these localized Wannier functions, and finally using the Wannier Tools package \cite{Wu2018} to calculate the Green's function spectrum.\cite{Sancho1985} The VASP simulations input into Wannier90 were simulated with noncollinear spins on a non-symmetrized $k$-mesh with and without SOC. The Mn magnetic moments in these simulations were constrained to point along [001]. Fine $k$-mesh mapping of the Weyl node locations in the 3D Brillouin zone was also performed on this tight-binding model. Berry curvature was determined using the Wannier90 \textsc{kslice} module. The band structure of Fig.\,\ref{fig:Figure5} was unfolded using BandUP.\cite{Medeiros2014,Medeiros2015} Orbital projections, density of states, and crystal orbital Hamiltonian populations for the high temperature phase were determined using LOBSTER.\cite{Dronskowski1993,Deringer2011,Maintz2013,Maintz2016} The density of states and orbital projections for the low temperature structure are reported using default VASP projections as LOBSTER does not support noncollinear magnetism. A post-process Gaussian smoothing with standard deviation 0.2\,eV was applied to all calculated density of states and COHPs. Structures are visualized with VESTA.\cite{Momma2011}

\subsection{Experimental}

Samples of Mn$_3$ZnC were produced by a two-step solid-state synthesis, following a previous report,\cite{Kaneko1987a} 
starting from stoichiometric quantities of Mn (Fisher Scientific, 99.95\%), Zn (Strem Chemicals, 99.9\%), and 
C (Alfa Aesar, 99\%) and adding 7 mass \% Zn during the second mixing step. Additional Zn was included to decrease the Mn:Zn 
ratio of the final material towards 3. Wavelength-dispersive X-ray fluorescence measurements performed on a 
Rigaku ZSX Primus IV suggest that the composition is close to Mn$_{3.16}$Zn$_{0.84}$C. Magnetic measurements were 
performed on a Quantum Design MPMS3 operating under vibrating sample magnetometer mode. High temperature measurements 
($T>$400\,K) including the PM-FM transition utilized the oven heater stick attachment option while lower temperature 
measurements used a standard brass sample holder. The magnetization curve in Fig.\,\ref{fig:Figure1} was generated by 
measuring magnetization at fields of $1$\,T, $0.1$\,T; below and above $300$\,K, respectively, and normalizing the high 
temperature magnetization curve such that there was no discontinuity at $300$\,K. $\partial M/\partial T$ and 
$\Delta S_M$ curves are calculated using methods reported previously.\cite{Bocarsly2018} Additional experimental details, 
including our attempts to develop an original microwave synthesis route for this and related antiperovskite carbides, are 
provided in the supporting material.

\section{Results}

\begin{figure}
\includegraphics[width=0.4\textwidth]{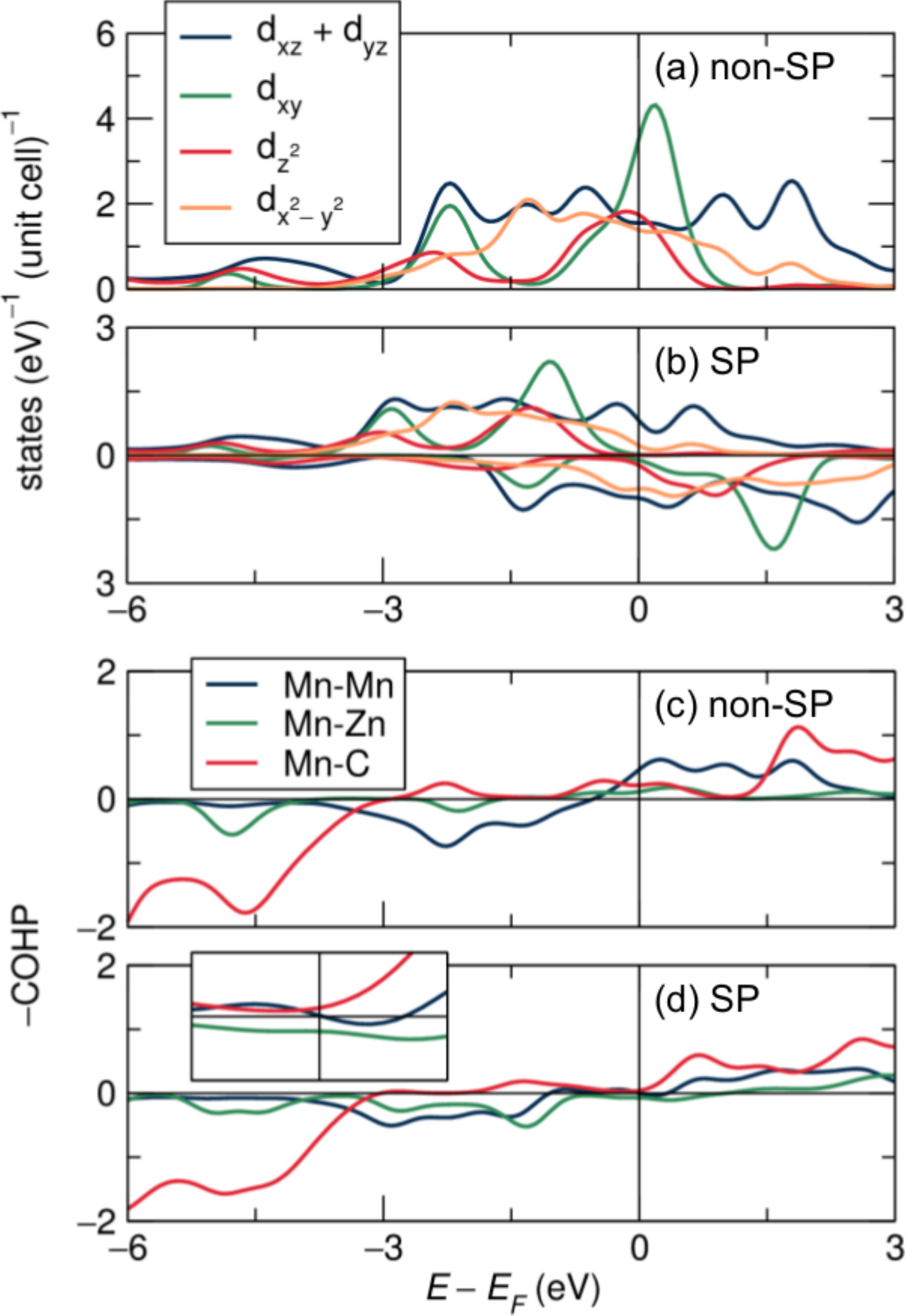}
\caption{\label{fig:Figure2} Density of states and crystal orbital Hamiltonian population for the ferromagnetic phase. Partial density of states of the dominant Mn $d$-orbital contributions are compared (a) without and (b) with spin polarization. The crystal orbital Hamiltonian populations for Mn-Mn, Mn-Zn, and Mn-C bonding are presented (c) without and (d) with spin polarization. Inset of (d) shows close-up of COHP in the region just about the Fermi level ($E_F\pm$0.5\,eV).}
\end{figure}

We first consider the density of states (DOS) for the cubic, ferromagnetic phase of Mn$_3$ZnC, as shown in Fig.\,\ref{fig:Figure2}. We consider only Mn $d$ states because other orbital contributions are small near the Fermi level (see supporting material). From Fig.\,\ref{fig:Figure2} (a), we can see that the DOS is relatively large at the Fermi level and that there is a large DOS peak just above this level. This is consistent with the hypothesized Stoner exchange mechanism for the ferromagnetism in this material. Indeed, in the spin-polarized calculation, Fig.\,\ref{fig:Figure2} (b), we find that the onset of ferromagnetism eliminates this DOS peak, reducing the Fermi-level DOS. The mechanism appears to be somewhat distinct from that originally proposed by Jardin and Labb{\'e}, however, as the near-Fermi level DOS peak in the non spin-polarized calculation originates from Mn $d_{xy}$ states, rather than $d_{xz}/d_{yz}$ states. Here, we consider the local symmetry of the Mn $d$ orbitals in terms of the cell coordinates (this is the preference in prior literature). From this perspective, we find that the $d_{xz}/d_{yz}$ orbitals are equivalent and point from one Mn towards its Mn neighbors and are involved in strong Mn-Mn bonding and some Mn-C bonding. $d_{z^2}$ and $d_{xy}$ orbitals have lobes pointing from Mn towards C and Zn atoms and are expected to be important in Mn-C and Mn-Zn bonding, respectively. Finally, $d_{x^2-y^2}$ orbitals point directly towards no other atom and are expected to be more weakly involved in bonding.

The proposed picture of a simple ferromagnetic distortion is further complicated when we consider these bonding interactions via the crystal orbital Hamiltonian population (COHP), a quantity derived from wave function overlap that is negative for bonding  and positive for antibonding interactions. In a normal ferromagnet, such as iron, the COHP shows significant antibonding at the Fermi level.\cite{Dronskowski2002} By splitting the spin populations, the normal ferromagnet is able to stabilize and fill a greater number of bonding states near the Fermi level, and relatively few antibonding states. The $d_{xy}$ orbitals, however, are not expected to be strongly (anti-)bonding; the COHP, Fig.\,\ref{fig:Figure2} (c), confirms the relatively weak strength of Mn-Zn interactions in this energy region. Rather than elimination of a sharp antibonding peak due to the $d_{xy}$ states, the spin-polarization appears to be stabilized by two effects: first, the reduction of the antibonding states just below the Fermi level and, second, the development of Mn-Zn bonding states originating from the originally weakly-bonding $d_{xy}$ band, Fig.\,\ref{fig:Figure2} (d). In the end, rather than a magnetic metal with a large number of bonding states at the Fermi level, we are left with a semimetal that has non-bonding character near $E_F$. Examining the COHP immediately about the Fermi level (Fig.\,\ref{fig:Figure2} (d), inset), we find that the Mn-Zn interactions are bonding and the Mn-C interactions are anti-bonding. The Mn-Mn COHP displays a subtle cross-over from antibonding to bonding at the Fermi level. Such a Fermi level COHP cross-over is typical of antiferromagnetic metals, although it usually involves bonding states below the Fermi level and antibonding states immediately above. In analog to the simple tight-binding chain model of 1$s$ orbitals, which forms a metal with a completely filled bonding band that can support a symmetry breaking lattice displacement of the atoms with a resulting bandgap, the Peierls distortion,\cite{Hoffmann1987} the bonding in Mn$_3$ZnC appears susceptible to symmetry breaking via both physical lattice distortion and antiferromagnetic ordering, either of which could distinguish the Mn sites and generate an electronic gap.

\begin{figure*}
\includegraphics[width=0.9\textwidth]{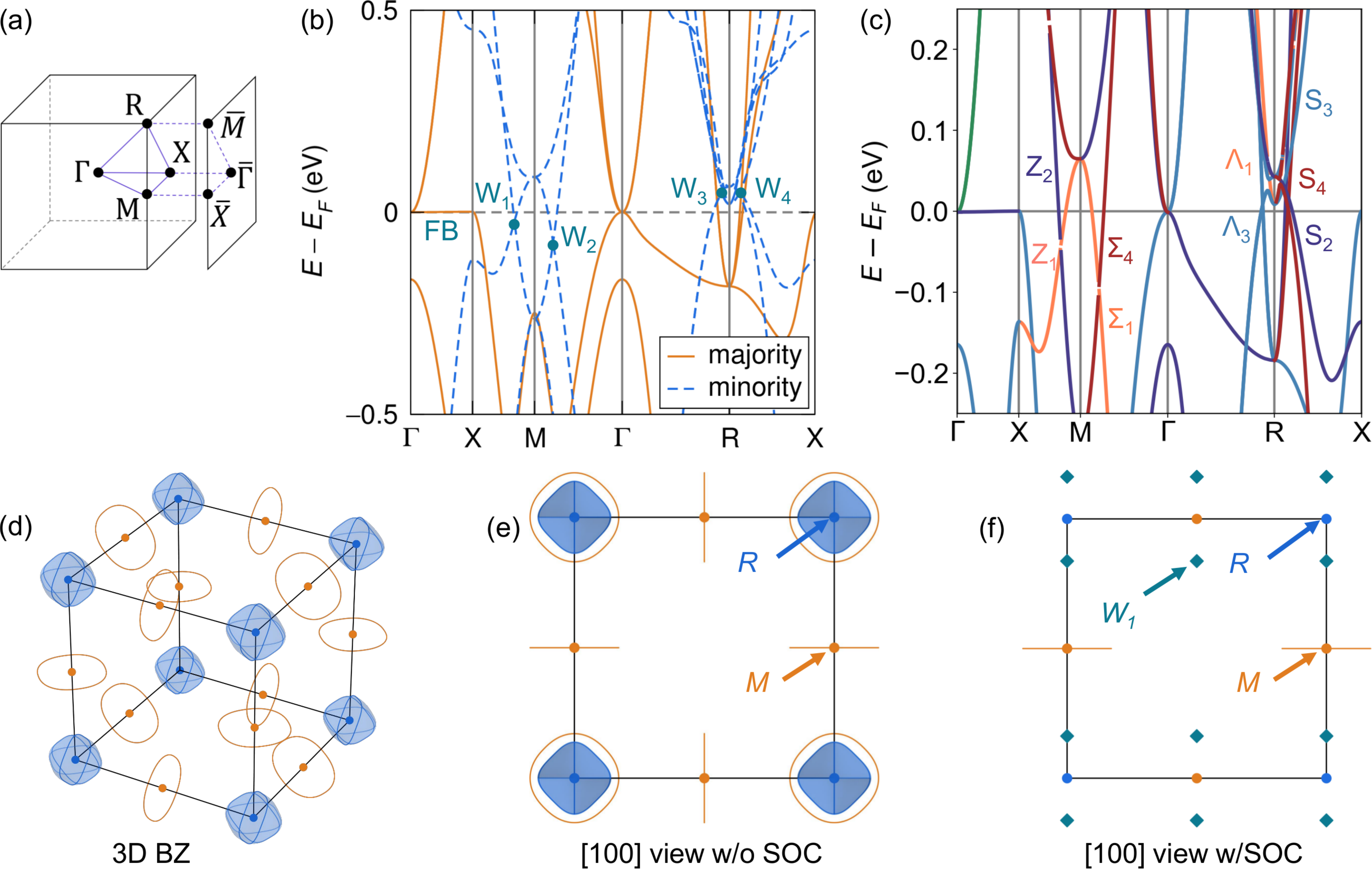}
\caption{\label{fig:Figure3} Bulk band structure and Weyl crossings. (a) high-symmetry points in the bulk Brillouin zone and (100)/(001) surface-projected Brillouin zone. (b) spin-polarized band structure showing a flat-band in the majority states and four types of near-Fermi level Weyl crossing. (c) close-up band structure with bands colored by irreducible representation. (d-f): schematics of Weyl surfaces in the 3D Brillouin zone. (d) 3D view without SOC. (e) (100) view without SOC. (f) (100) view with SOC. In the absence of SOC, Nodal loops are displayed around $M$ (orange) and 3D nodal surfaces around $R$ (blue). With magnetization along [001], SOC gaps the nodal surfaces and many of the nodal loops, leaving behind $k_z=0$ nodal loops and isolated $W_1$ nodes.}
\end{figure*}

This ferromagnetic semimetal phase becomes much more interesting when we consider the electronic band structure in Fig.\,\ref{fig:Figure3}, which reveals several Weyl nodes. Fig.\,\ref{fig:Figure3} (a) depicts the bulk cubic Brillouin zone. Fig.\,\ref{fig:Figure3} (b) presents the spin-polarized band structure with orange majority bands and blue minority bands. Five features of interest are labeled. First, despite the gapping of the density of states at the Fermi level, a flat band, FB, persists along $\Gamma$-$X$ right at the Fermi level. Along $X$-$M$ and $M$-$\Gamma$, we find linear band crossings, Weyl nodes, in the minority bands, labeled $W_1$ and $W_2$. On either side of the $R$ point, we also find linear band crossings just above the Fermi level involving both the majority and minority bands. The Weyl nodes on either side of $R$ are labeled $W_3$ and $W_4$. Note, however, that there are actually several Weyl nodes on either side of $R$ and this labeling is a reduction for the sake of simplicity.

The colors of the bands in Fig.\,\ref{fig:Figure3} (c) depict the irreducible representations (irreps) of bands involved in the Weyl crossings. A band crossing is protected against gapping if the irreps of the two crossing bands are different; this represents orthogonality of the electronic states generating these bands. In the absence of spin-orbit coupling (SOC), we can consider this system in the paramagnetic space group ($Pm\overline{3}m$, \#221). The point group of the $k$-vectors along $X$-$M$, $M$-$\Gamma$ and $R$-$X$ is $C_{2v}$ and along $\Gamma$-$R$ the point group is $C_{3v}$. The $Z_1$ and $Z_2$ irreps differ in their symmetry with respect to $C_2$ rotation and the mirror operation $\sigma_v^\prime$, indicating that the $W_1$ node is protected by these operations. The $\Sigma_1$ and $\Sigma_4$ irreps differ with respect to $C_2$ rotation and the mirror operation $\sigma_v$. The $W_4$ nodes involve crossings of $S_2,S_3,S_4$ bands. The nearest Weyl crossings to the Fermi level along $\Gamma$-$R$ and $R$-$X$ actually involve bands with the same irrep, $\Lambda_3$ and $S_4$, respectively. However, these crossings have bands of orthogonal spinor character, as can be seen when comparing with Fig.\,\ref{fig:Figure3} (b). When SOC is incorporated in the simulation, we must consider the coupling of the magnetic moment to the lattice as well as majority / minority spin mixing interactions. Spin mixing gaps the lowest energy $W_3,W_4$ crossings and the breaking of $C_2$ and mirror symmetries along $R$-$X$ gaps the $W_4$ Weyl nodes. The $W_3$ Weyl crossings due to $\Lambda_1,\Lambda_3$ crossings, meanwhile, are left untouched by SOC. When considering the $k_x=0$ Brillouin zone plane, which in the non-SOC calculation had been equivalent, however, both nodes are not preserved. Viewing the $W_1$ and $W_2$ nodes, on the $k_x=0$ plane, we find that two of the four $W_1$ nodes and all four of the $W_2$ nodes gap out (see supplement for additional details on the effect of SOC). In all cases, the magnitude of the SOC gaps in this material are relatively small due to its low-mass $3d$ electrons.

Mapping the Weyl nodes in the 3D Brillouin zone, we find that the Weyl nodes in this system are not isolated points, but instead nodal lines and surfaces as depicted in Fig.\,\ref{fig:Figure3} (d,e). The $W_1$ and $W_2$ nodes are part of a nodal line about the $M$ point whereas the lowest energy $W_3$ and $W_4$ nodes form a 3D connected surface about the $R$ point. Because of the large number of $W_{3,4}$ Weyl crossings, additional $W_{3,4}$ nodal surfaces may exist which have not been considered. In general, although individual band crossings and nodal lines can be protected, there are no symmetries that can fully protect a 3D nodal surface in the presence of SOC.\cite{Wang2018,Teicher2019a} The nodal surfaces depicted in Fig.\,\ref{fig:Figure3}(d,e) about $R$ are no exception. Because these surfaces result from the crossing of bands of opposite spinor character, and SOC allows spin-channel mixing, these nodal surfaces gap out and the nodal surface is no longer realized in the final electronic structure after SOC has been incorporated. We find that some of the $M$ nodal lines, however, are protected. The nodal loops on the $k_z=0$ plane are protected against gapping, while the nodal loops on the $k_x,k_y=0$ planes gap, leaving only isolated Weyl crossings on the $X$-$M$ lines. This results in the spin-orbit-gapped configuration shown in Fig.\,\ref{fig:Figure3} (f). 

The protection of the $M$ nodal lines is a direct result of the mirror symmetry in the cubic antiperovskite structure. While we have already shown that a combination of $C_2$ and mirror symmetries protects the band crossings along the $\Gamma$-$M$ and $M$-$X$ high symmetry lines, recomputing the irreps along a generic path through the loop with point group $C_s$ shows the band crossings to be protected by $\overline{C_2}$, which is equivalent to a mirror. The $M$ nodal lines on the $k_z=0$ plane are preserved under SOC because the [001]-oriented collinear magnetic moment does not break mirror symmetry on this plane. Likewise, the $M$ nodal lines on the $k_x,k_y=0$ planes gap because the magnetic moment does break the mirror symmetry on these planes---except at the $W_1$ nodes where the nodal line is tangent to the $k_z=0$ plane and perpendicular to the [001]-magnetic moment.

These nodal lines are similar to those described in a theoretical tight-binding model of Ca$_3$P$_2$ with imposed spin and previously predicted in alloys of the real material ZrCo$_2$Sn (also under [001] magnetization).\cite{Schnyder2016,Wang2016} Due to the soft ferromagnetism and readily-reorientable magnetization in ZrCo$_2$Sn, it was proposed that the effective number of Weyl nodes can be tuned by changing the applied field direction. Mn$_3$ZnC also appears to be a soft ferromagnet and could exhibit similar tunability.

\begin{figure*}
\includegraphics[width=1.0\textwidth]{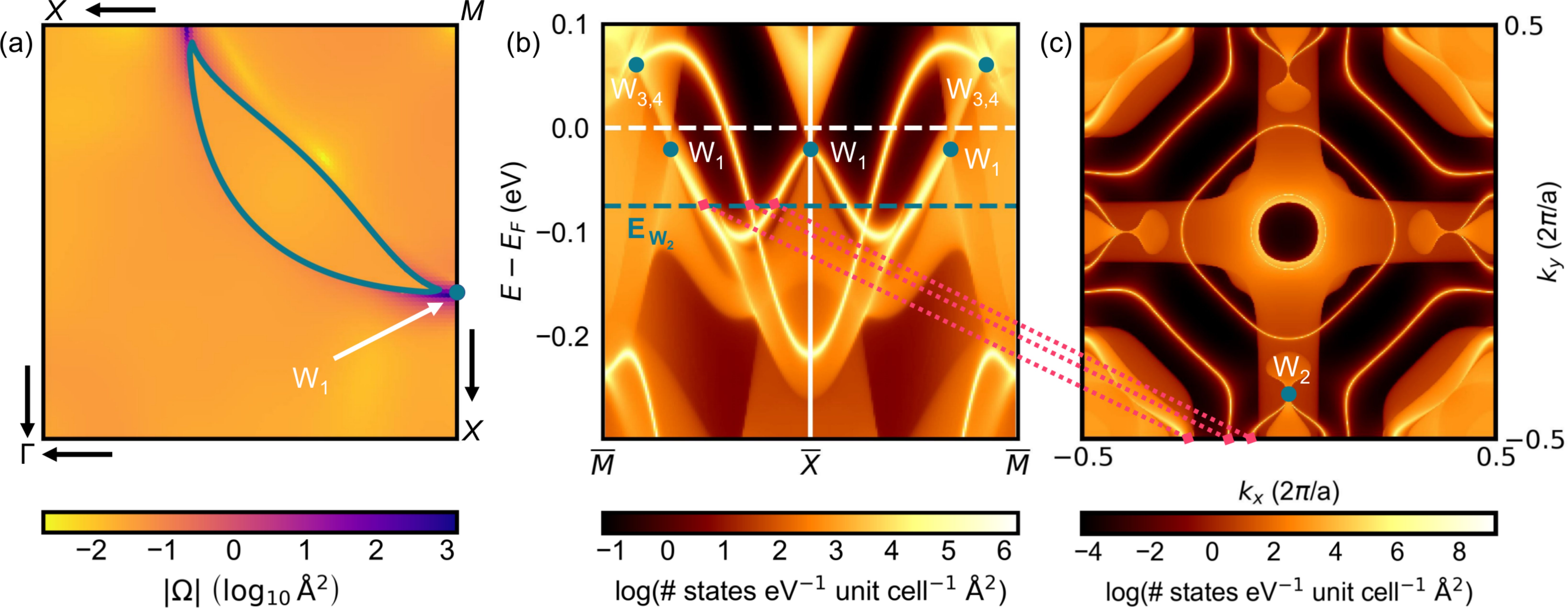}
\caption{\label{fig:Figure4} Characterization of Weyl behavior. (a) $k_z = 0$ cut through the bulk BZ showing large Berry curvature near the $W_1$ Weyl crossings. A constant energy band-cross-section is overlaid in teal. (b) surface density of states calculation showing brightly-colored surface states connecting between $W_{3,4}$ and between $W_1$ Weyl points. The energy level of the $W_2$ Weyl points is also shown. (c) constant energy Fermi arc calculation for the (001) surface at the energy level of the $W_2$ nodes. Dotted pink lines connect between surface states at the $W_2$ level in (b) and the locations of these states in (c) to guide the eye.}
\end{figure*}

Now that the existence of a protected $k_z=0$ nodal loop has been established, we might expect to see drumhead surface states connecting between the nodes of this loop on the (001) surface of the material. Fig.\,\ref{fig:Figure4} characterizes the Weyl nodes and their surface states. Weyl nodes can be described as sources and sinks of a quantity called the Berry curvature or Berry flux, defined as:
$$ \nabla_\mathbf{k} \times \braket{u_\mathbf{k}|i\partial_\mathbf{k}u_\mathbf{k}} $$
where $u_\mathbf{k}$ represents the Bloch wave function.\cite{Vanderbilt2018} Low energy electronic excitations analogous to quantum Hall surface states, Fermi arcs, are topologically protected between Weyl nodes emitting and collecting Berry flux.\cite{Armitage2018} Fig.\,\ref{fig:Figure4} (a) shows the magnitude of the Berry flux in the $k_z=0$ plane at an energy level near the $W_1$ nodes, highlighting large Berry flux concentration at the nodes. Fig.\,\ref{fig:Figure4} (b) and (c) present the Weyl surface states projected on a (001) surface. We can see bright surface bands connecting between two $W_1$ nodes and between two $W_{3,4}$ nodes (Note, following Fig \ref{fig:Figure3} (a), that the $R$-$M$ and $X$-$M$ lines are projected onto the $\overline{M}$ and $\overline{X}$ points when flattening the cubic Brillouin zone along $k_z$). The energy level of the $W_2$ nodes is overlaid in teal. In Fig.\,\ref{fig:Figure4} (c), we can see a Fermi surface cut taken at the $W_2$ energy level with $W_2$ nodes clearly visible along $\overline{\Gamma}$-$\overline{X}$. Fermi arcs emitting from the $W_2$ points can be seen to connect to $W_2$ nodes in the neighboring Brillouin zone due to the periodic boundary conditions. The $\overline{M}$-$\overline{X}$-$\overline{M}$ line in Fig.\,\ref{fig:Figure4} (b) corresponds to the edges of the plot in Fig.\,\ref{fig:Figure4} (c). Comparing between Fig.\,\ref{fig:Figure4} (b) and (c), we see that the surface states connecting the $W_1$ nodes are identical to the surface states connecting the $W_2$ nodes; there is indeed a drumhead surface state connecting $W_1$, $W_2$ and the other nodes on the nodal line about the $M$ point. 

In general, materials with electronic structure features like those in Fig.\,\ref{fig:Figure3} (b)---degenerate flat bands and near-Fermi level band crossings---and especially semimetals with a large concentration of Weyl crossings like this nodal-loop compound, often distort at low temperatures. The energy-lowering transitions that tend to break such nodes are frequently referenced to the idealized models of Peierls or Jahn-Teller distortions. The well-established experimental fact is that a symmetry-breaking low-temperature distortion does activate in this material, involving both a tetragonal stretch of the atomic positions and a magnetic reorientation of the spins with partial antiferromagnetic coupling (Fig.\,\ref{fig:Figure1}).

\begin{figure}
\includegraphics[width=0.45\textwidth]{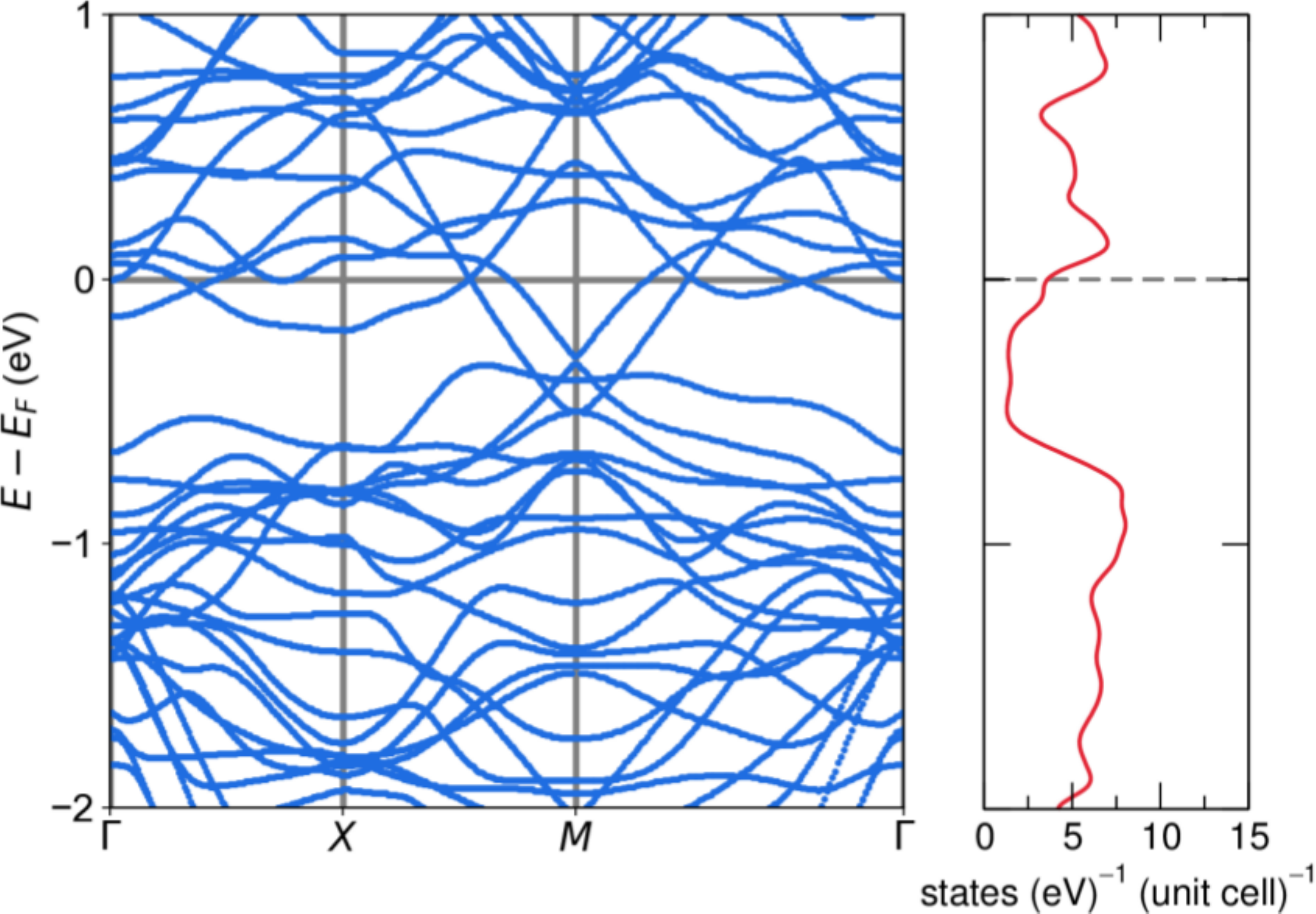}
\caption{\label{fig:Figure5} Band structure and density of states in the non-collinear, low-temperature phase. (Total) DOS values have been normalized to the high-temperature unit cell to provide fair comparison to Fig.\,\ref{fig:Figure2}. A large gap is seen that can be accessed by decreasing the electron count through slight electron deficiency. This unfolded band structure allows a direct comparison of the ferrimagnetic tetragonal cell bands to the band structure of the ferromagnetic cubic unit cell. Although there is significant gapping and removal of Weyl nodes just below the Fermi level, a small number of Weyl nodes persist at $E_F$.}
\end{figure}

The density of states of the low temperature structure provide valuable insight for the origin of the low temperature phase. Fig.\,\ref{fig:Figure5} shows the DOS and band structure of the tetragonal low temperature structure with fully non-collinear spins and associated antiferromagnetic coupling. A large pseudogap is seen to open up in this case, as would be expected from the description of the antiferromagnetic ordering as a Peierls-like symmetry breaking. We also considered simulations with the tetragonal low temperature structure with no spin polarization and collinear, ferromagnetic polarization. However, because the tetragonal distortion is fairly small, the density of states for these phases exhibit only minor changes with respect to those of Fig.\,\ref{fig:Figure2}. Further, when allowed to relax, the non-spin polarized and collinear spin polarized structures relax into the cubic structure. This analysis suggests that though both are coupled, it is probably the new magnetic ordering, rather than tetragonal distortion alone, which drives the transition.

The relationship between the new magnetic ordering and potential electronic instabilities, flat bands and Weyl nodes, can be examined via the band structure of the low temperature phase in Fig.\,\ref{fig:Figure5}. In order to provide a direct comparison to the band structure and Weyl nodes of the ferromagnetic phase in Fig.\,\ref{fig:Figure3}, we have unfolded the bands of the low temperature cell into a Brillouin zone corresponding to that of the low temperature cubic primitive cell. We see that, in the low temperature structure, the flat band and several of the Weyl nodes have disappeared. The DOS pseudogap can be seen in the relatively empty region of the band structure about 0.2\,eV below the Fermi level. Despite the formation of the pseudogap, a few Weyl nodes remain at the calculated Fermi level. Part of the discrepancy between our simulation and the expectation for the stabilization of a pseudogap \textit{at the Fermi level} likely results from electron deficiency in the real material. As early as Butters and Myers' first study, Mn$_3$ZnC was found to have varying composition and magnetic properties with nominal Mn:Zn ratios greater than $3$.\cite{Butters1955a} The Mn:Zn ratio of our best sample, measured by wavelength-dispersive X-ray fluorescence, was close to 3.76, which, in a rigid band approximation, would result in a deficiency of 0.8 electrons per primitive cubic cell and a Fermi level near the exact center of the pseudogap. We expect that the true Mn:Zn ratio in our samples lies somewhere between the idealized value of 3.00 and the measured value of 3.76; while samples of this material appear pure in laboratory X-ray diffraction, synchrotron X-ray diffraction suggests that small nonmagnetic impurities of carbides Mn$_5$C$_2$ and Mn$_7$C$_3$ as well as MnO (which orders antiferromagnetically, but far lower in temperature than the transitions discussed here) may be present in addition to the majority Mn$_3$ZnC phase.\cite{Karen1991}

\begin{figure*}
\includegraphics[width=0.9\textwidth]{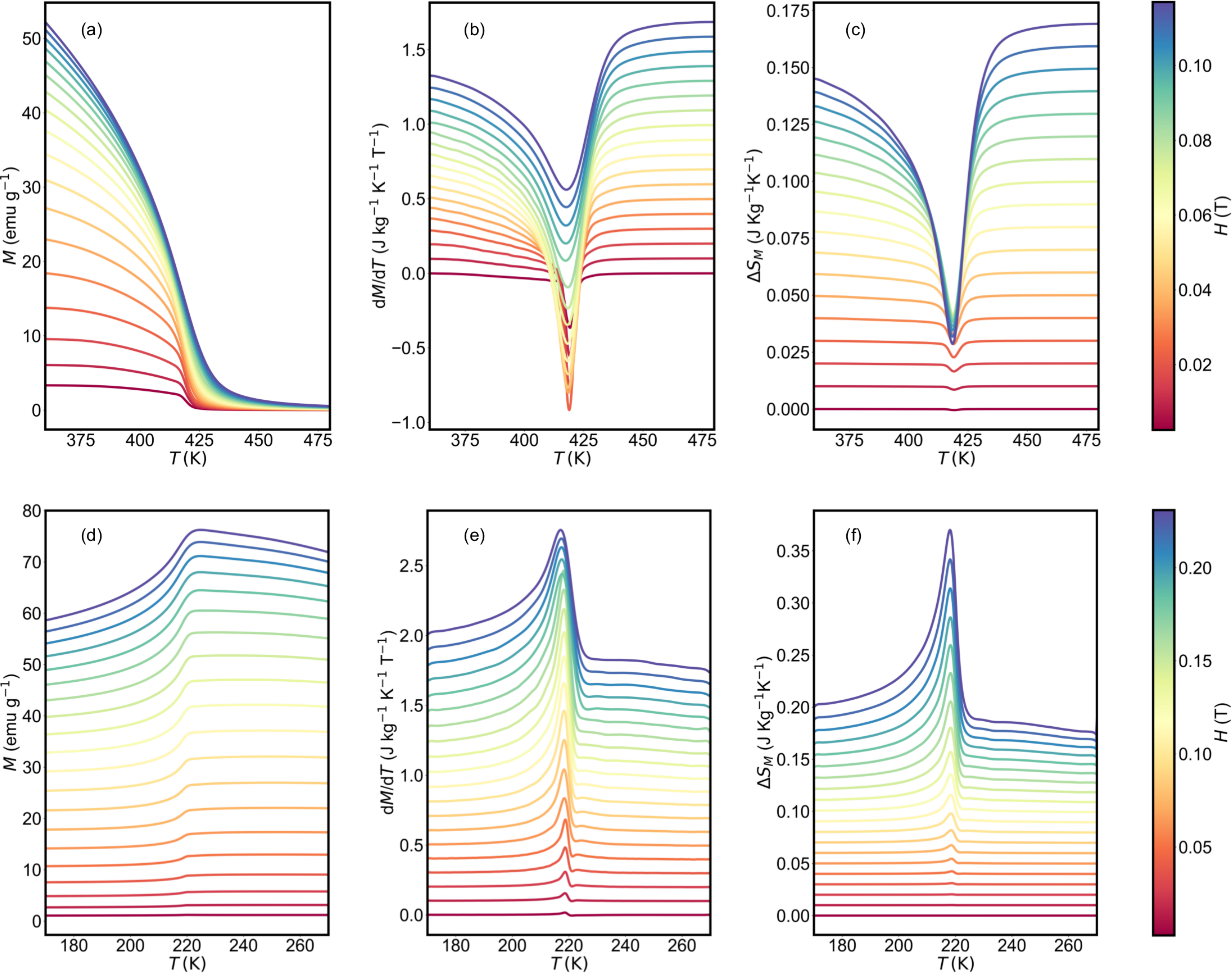}
\caption{\label{fig:Figure6} Magnetoentropic mapping of the PM-FM and FM-NCL transitions. (a) line plots of the magnetization, (b) magnetization derivative ($\partial M$/$\partial T$), and (c) magnetic entropy change ($\Delta S_M$) taken under varying magnetic field are shown across the paramagnetic to ferromagnetic transition that occurs at approximately $420$\,K in our sample. (d-f) provide similar plots for the FM-NCL transition ($\approx219$\,K). Lines in (b,e) and (c,f) are offset by 0.1 and 0.01 units, respectively, for visual clarity.}
\end{figure*}

Motivated by interest in these electronic structure changes associated with magnetostructurally-coupled transitions, we performed experimental magnetoentropic characterization of both the high and low temperature transitions. Fig.\,\ref{fig:Figure6} (a) presents magnetization data measured across the paramagnetic to ferromagnetic transition, which takes place near $420$\,K in our sample. While magnetization has been measured in older studies, to the best of our knowledge no studies of the the sharp, low field magnetization curves for either of the two magnetic transitions have been previously reported. Fig.\,\ref{fig:Figure6} (b) shows the partial derivative, $\partial M$/$\partial T$, of these magnetization curves, displaying a large transition peak at low fields that reduces in magnitude and broadens as the field is increased. Fig.\,\ref{fig:Figure6} (c) shows the magnetic entropy change, $\Delta S_M$ across the transition. The magnitude of $\Delta S_M$ is a direct probe of the magnetostructural coupling strength in this compound.\cite{Franco2012} The peak $\Delta S_M$ value is significant but not anomalous, suggesting medium magnetostructural coupling strength in this compound. Negative $\Delta S_M$ transitions are typical of the magnetocaloric effect seen in paramagnetic to ferromagnetic transitions in many magnetic materials. Characterization of the antiferromagnetic ordering transition in (d-f) is consistent with this finding. This transition is qualitatively similar in that it is sharp for low applied field, broad at high field and has $\Delta S_M$ values with approximately the same magnitude of $\approx0.1$ J Kg$^{-1}$ K$^{-1}$. The positive sign of $\Delta S_M$ is typical of the inverse magnetocaloric effect seen at many antiferromagnetic ordering transitions.

\section{Conclusions}

We have shown electronic structure simulations predicting that the room temperature phase of Mn$_3$ZnC is an exotic Weyl nodal line semimetal with nodal loops, isolated Weyl nodes, and drumhead surface states. The magnetic and electronic characterizations of the two transitions in this material that create and destroy this phase, meanwhile, appear relatively conventional. The upper transition can be explained by the reduction of near-Fermi level antibonding states and strengthening of Mn-Zn bonding, while the lower transition can be explained by the need to break symmetry and open a gap by expanding the unit cell to allow for antiferromagnetic ordering. An electronic Peierls distortion through antiferromagnetic ordering can allow for spin population energy shifts and the formation of a pseudogap that remove flatbands and Weyl nodes near the Fermi level. Despite the significant pseudogap, we found that a limited number of Weyl nodes still persist near the Fermi level in the low-temperature structure. However, even a small electron deficiency, expected based on the tendency towards Zn deficiency in experimental work, moves the Fermi level into the pseudogap of the low temperature structure.

There is interest at present in the Peierls-like structural distortions and phonon resonances associated with Weyl nodes.\cite{Park2019} The finding of Weyl nodal lines in Mn$_3$ZnC, a classic magnetic transition material, suggests that Weyl instability may play a role in a much wider range of magnetostructurally-coupled materials. Many other magnetic materials that have a low temperature antiferromagnetic ordering likely transition through a semimetal state with near-Fermi level Weyl nodes. In fact, prototypical itinerant antiferromagnet chromium itself, the material for which the electronic Peierls instability concept was coined, hosts Fermi level Dirac crossings in non-spin polarized calculations, some of which are preserved in its transition to a low temperature antiferromagnetic ordering.\cite{Decker2002} One major distinction between Mn$_3$ZnC and conventional antiferromagnetic materials is the reversed Peierls-like bonding structure with antibonding states below the Fermi level, non-bonding states at $E_F$, and bonding states just above. This reversed bonding structure is indicative of band inversion and could prove a useful hallmark in the search for topologically-interesting magnets.

In addition to the bonding analysis we have presented to explain the structural transitions, important future work will focus on rigorously disentangling the relationship between Weyl nodes, flat bands, and phonon-mediated instabilities in Mn$_3$ZnC and related semimetals. There is incredible potential for nesting in the Fermi surface of Mn$_3$ZnC; the flat-bands alone, which perfectly bisect the Brillouin zone in $k_x,k_y,k_z$ directions, provide maximal nesting at the calculated Fermi level in any supercell scheme as well as the tantalizing prospect of coupling flat-band-related correlation effects to Weyl physics. We urge caution. Just as flat band degeneracies in DFT are not a guarantee of interesting correlation effects in experiment, substantial evidence suggests that calculations of Fermi surface nesting are insufficient to prescriptively predict charge density waves, spin density waves and other lattice incommensurate instabilities.\cite{Johannes2008} An important first step has been provided by a recent study on (TaSe$_4$)$_2$I, which relates the characteristic $q$-spacings of the Weyl nodes, peaks in the electronic susceptibility, and CDW-modulation vectors observed in experimental X-ray measurements.\cite{Shi2018}

Overall, our results suggest that compounds which display Weyl-like features in idealized high-symmetry structures may actually undergo transitions to more complex ground states than initially supposed.

\section{Supplementary Material}
See supplementary material for additional experimental and computational details including discussion of the effects of spin-orbit coupling on the electronic structure.

\section{Acknowledgments}

The work at UC Santa Barbara was supported by the National Science Foundation (NSF) through DMR 1710638.
LMS has been supported by the Princeton Center for Complex Materials, a Materials Research Science and 
Engineering Center (MRSEC) DMR 1420541. The research reported here made use of shared facilities of the 
NSF MRSEC at UC Santa Barbara, DMR 1720256. The UC Santa Barbara MRSEC is a member of the Materials Research 
Facilities Network (www.mrfn.org). We also acknowledge the use of the computing facilities of the Center 
for Scientific Computing at UC Santa Barbara supported by NSF CNS 1725797 and NSF DMR 1720256. Use of the 
Advanced Photon Source at Argonne National Laboratory was supported by the U.S. Department of Energy, Office 
of Science, Office of Basic Energy Sciences, under Contract No. DE-AC02-06CH11357. IKS gratefully acknowledges support from
the IRES: Cooperative for Advanced Materials in Energy-Related Applications (NSF-OISE 1827034) 
and from AoA Materials Science, Chalmers University of Technology. SMLT has been supported by the National 
Science Foundation Graduate Research Fellowship Program under Grant No. DGE-1650114. Any opinions, findings, 
and conclusions or recommendations expressed in this material are those of the authors and do not necessarily 
reflect the views of the National Science Foundation. 

\section*{References}
\bibliography{my_bib}

\begin{thebibliography}{62}%
\makeatletter
\providecommand \@ifxundefined [1]{%
 \@ifx{#1\undefined}
}%
\providecommand \@ifnum [1]{%
 \ifnum #1\expandafter \@firstoftwo
 \else \expandafter \@secondoftwo
 \fi
}%
\providecommand \@ifx [1]{%
 \ifx #1\expandafter \@firstoftwo
 \else \expandafter \@secondoftwo
 \fi
}%
\providecommand \natexlab [1]{#1}%
\providecommand \enquote  [1]{``#1''}%
\providecommand \bibnamefont  [1]{#1}%
\providecommand \bibfnamefont [1]{#1}%
\providecommand \citenamefont [1]{#1}%
\providecommand \href@noop [0]{\@secondoftwo}%
\providecommand \href [0]{\begingroup \@sanitize@url \@href}%
\providecommand \@href[1]{\@@startlink{#1}\@@href}%
\providecommand \@@href[1]{\endgroup#1\@@endlink}%
\providecommand \@sanitize@url [0]{\catcode `\\12\catcode `\$12\catcode
  `\&12\catcode `\#12\catcode `\^12\catcode `\_12\catcode `\%12\relax}%
\providecommand \@@startlink[1]{}%
\providecommand \@@endlink[0]{}%
\providecommand \url  [0]{\begingroup\@sanitize@url \@url }%
\providecommand \@url [1]{\endgroup\@href {#1}{\urlprefix }}%
\providecommand \urlprefix  [0]{URL }%
\providecommand \Eprint [0]{\href }%
\providecommand \doibase [0]{http://dx.doi.org/}%
\providecommand \selectlanguage [0]{\@gobble}%
\providecommand \bibinfo  [0]{\@secondoftwo}%
\providecommand \bibfield  [0]{\@secondoftwo}%
\providecommand \translation [1]{[#1]}%
\providecommand \BibitemOpen [0]{}%
\providecommand \bibitemStop [0]{}%
\providecommand \bibitemNoStop [0]{.\EOS\space}%
\providecommand \EOS [0]{\spacefactor3000\relax}%
\providecommand \BibitemShut  [1]{\csname bibitem#1\endcsname}%
\let\auto@bib@innerbib\@empty
\bibitem [{\citenamefont {Barsoum}(2000)}]{Barsoum2000}%
  \BibitemOpen
  \bibfield  {author} {\bibinfo {author} {\bibfnamefont {M.~W.}\ \bibnamefont
  {Barsoum}},\ }\href {\doibase 10.1016/S0079-6786(00)00006-6} {\bibfield
  {journal} {\bibinfo  {journal} {Prog. Solid State Ch.}\ }\textbf {\bibinfo
  {volume} {28}},\ \bibinfo {pages} {201} (\bibinfo {year} {2000})}\BibitemShut
  {NoStop}%
\bibitem [{\citenamefont {Kimura}, \citenamefont {Mishima},\ and\ \citenamefont
  {Liu}(2001)}]{Kimura2001}%
  \BibitemOpen
  \bibfield  {author} {\bibinfo {author} {\bibfnamefont {Y.}~\bibnamefont
  {Kimura}}, \bibinfo {author} {\bibfnamefont {Y.}~\bibnamefont {Mishima}}, \
  and\ \bibinfo {author} {\bibfnamefont {C.}~\bibnamefont {Liu}},\ }\href
  {\doibase 10.1016/S0966-9795(01)00113-3} {\bibfield  {journal} {\bibinfo
  {journal} {Intermetallics}\ }\textbf {\bibinfo {volume} {9}},\ \bibinfo
  {pages} {1069} (\bibinfo {year} {2001})}\BibitemShut {NoStop}%
\bibitem [{\citenamefont {Wen}\ \emph {et~al.}(2009)\citenamefont {Wen},
  \citenamefont {Wang}, \citenamefont {Sun}, \citenamefont {Nie}, \citenamefont
  {Fang},\ and\ \citenamefont {Tian}}]{Wen2009}%
  \BibitemOpen
  \bibfield  {author} {\bibinfo {author} {\bibfnamefont {Y.}~\bibnamefont
  {Wen}}, \bibinfo {author} {\bibfnamefont {C.}~\bibnamefont {Wang}}, \bibinfo
  {author} {\bibfnamefont {Y.}~\bibnamefont {Sun}}, \bibinfo {author}
  {\bibfnamefont {M.}~\bibnamefont {Nie}}, \bibinfo {author} {\bibfnamefont
  {L.}~\bibnamefont {Fang}}, \ and\ \bibinfo {author} {\bibfnamefont
  {Y.}~\bibnamefont {Tian}},\ }\href {\doibase 10.1016/j.ssc.2009.06.009}
  {\bibfield  {journal} {\bibinfo  {journal} {Solid State Commun.}\ }\textbf
  {\bibinfo {volume} {149}},\ \bibinfo {pages} {1519 } (\bibinfo {year}
  {2009})}\BibitemShut {NoStop}%
\bibitem [{\citenamefont {Hamada}\ and\ \citenamefont
  {Takenaka}(2011)}]{Hamada2011}%
  \BibitemOpen
  \bibfield  {author} {\bibinfo {author} {\bibfnamefont {T.}~\bibnamefont
  {Hamada}}\ and\ \bibinfo {author} {\bibfnamefont {K.}~\bibnamefont
  {Takenaka}},\ }\href {\doibase 10.1063/1.3540604} {\bibfield  {journal}
  {\bibinfo  {journal} {J. Appl. Phys.}\ }\textbf {\bibinfo {volume} {109}},\
  \bibinfo {pages} {07E309} (\bibinfo {year} {2011})}\BibitemShut {NoStop}%
\bibitem [{\citenamefont {Kamishima}\ \emph {et~al.}(2000)\citenamefont
  {Kamishima}, \citenamefont {Goto}, \citenamefont {Nakagawa}, \citenamefont
  {Miura}, \citenamefont {Ohashi}, \citenamefont {Mori}, \citenamefont
  {Sasaki},\ and\ \citenamefont {Kanomata}}]{Kamishima2000}%
  \BibitemOpen
  \bibfield  {author} {\bibinfo {author} {\bibfnamefont {K.}~\bibnamefont
  {Kamishima}}, \bibinfo {author} {\bibfnamefont {T.}~\bibnamefont {Goto}},
  \bibinfo {author} {\bibfnamefont {H.}~\bibnamefont {Nakagawa}}, \bibinfo
  {author} {\bibfnamefont {N.}~\bibnamefont {Miura}}, \bibinfo {author}
  {\bibfnamefont {M.}~\bibnamefont {Ohashi}}, \bibinfo {author} {\bibfnamefont
  {N.}~\bibnamefont {Mori}}, \bibinfo {author} {\bibfnamefont {T.}~\bibnamefont
  {Sasaki}}, \ and\ \bibinfo {author} {\bibfnamefont {T.}~\bibnamefont
  {Kanomata}},\ }\href {\doibase 10.1103/PhysRevB.63.024426} {\bibfield
  {journal} {\bibinfo  {journal} {Phys. Rev. B}\ }\textbf {\bibinfo {volume}
  {63}},\ \bibinfo {pages} {024426} (\bibinfo {year} {2000})}\BibitemShut
  {NoStop}%
\bibitem [{\citenamefont {He}\ \emph {et~al.}(2001)\citenamefont {He},
  \citenamefont {Huang}, \citenamefont {Ramirez}, \citenamefont {Wang},
  \citenamefont {Regan}, \citenamefont {Rogado}, \citenamefont {Hayward},
  \citenamefont {Haas}, \citenamefont {Slusky}, \citenamefont {Inumara},
  \citenamefont {Zandbergen}, \citenamefont {Ong},\ and\ \citenamefont
  {Cava}}]{He2001}%
  \BibitemOpen
  \bibfield  {author} {\bibinfo {author} {\bibfnamefont {T.}~\bibnamefont
  {He}}, \bibinfo {author} {\bibfnamefont {Q.}~\bibnamefont {Huang}}, \bibinfo
  {author} {\bibfnamefont {A.~P.}\ \bibnamefont {Ramirez}}, \bibinfo {author}
  {\bibfnamefont {Y.}~\bibnamefont {Wang}}, \bibinfo {author} {\bibfnamefont
  {K.~A.}\ \bibnamefont {Regan}}, \bibinfo {author} {\bibfnamefont
  {N.}~\bibnamefont {Rogado}}, \bibinfo {author} {\bibfnamefont {M.~A.}\
  \bibnamefont {Hayward}}, \bibinfo {author} {\bibfnamefont {M.~K.}\
  \bibnamefont {Haas}}, \bibinfo {author} {\bibfnamefont {J.~S.}\ \bibnamefont
  {Slusky}}, \bibinfo {author} {\bibfnamefont {K.}~\bibnamefont {Inumara}},
  \bibinfo {author} {\bibfnamefont {H.~W.}\ \bibnamefont {Zandbergen}},
  \bibinfo {author} {\bibfnamefont {N.~P.}\ \bibnamefont {Ong}}, \ and\
  \bibinfo {author} {\bibfnamefont {R.~J.}\ \bibnamefont {Cava}},\ }\href
  {\doibase 10.1038/35075014} {\bibfield  {journal} {\bibinfo  {journal}
  {Nature}\ }\textbf {\bibinfo {volume} {411}},\ \bibinfo {pages} {54}
  (\bibinfo {year} {2001})}\BibitemShut {NoStop}%
\bibitem [{\citenamefont {Mollah}(2004)}]{Mollah2004}%
  \BibitemOpen
  \bibfield  {author} {\bibinfo {author} {\bibfnamefont {S.}~\bibnamefont
  {Mollah}},\ }\href {\doibase 10.1088/0953-8984/16/43/r01} {\bibfield
  {journal} {\bibinfo  {journal} {J. Phys.: Condens. Matter}\ }\textbf
  {\bibinfo {volume} {16}},\ \bibinfo {pages} {R1237} (\bibinfo {year}
  {2004})}\BibitemShut {NoStop}%
\bibitem [{\citenamefont {Kariyado}\ and\ \citenamefont
  {Ogata}(2011)}]{Kariyado2011}%
  \BibitemOpen
  \bibfield  {author} {\bibinfo {author} {\bibfnamefont {T.}~\bibnamefont
  {Kariyado}}\ and\ \bibinfo {author} {\bibfnamefont {M.}~\bibnamefont
  {Ogata}},\ }\href {\doibase 10.1143/JPSJ.80.083704} {\bibfield  {journal}
  {\bibinfo  {journal} {J. Phys. Soc. Jpn.}\ }\textbf {\bibinfo {volume}
  {80}},\ \bibinfo {pages} {083704} (\bibinfo {year} {2011})}\BibitemShut
  {NoStop}%
\bibitem [{\citenamefont {Kariyado}\ and\ \citenamefont
  {Ogata}(2012)}]{Kariyado2012}%
  \BibitemOpen
  \bibfield  {author} {\bibinfo {author} {\bibfnamefont {T.}~\bibnamefont
  {Kariyado}}\ and\ \bibinfo {author} {\bibfnamefont {M.}~\bibnamefont
  {Ogata}},\ }\href {\doibase 10.1143/JPSJ.81.064701} {\bibfield  {journal}
  {\bibinfo  {journal} {J. Phys. Soc. Jpn.}\ }\textbf {\bibinfo {volume}
  {81}},\ \bibinfo {pages} {064701} (\bibinfo {year} {2012})}\BibitemShut
  {NoStop}%
\bibitem [{\citenamefont {Castro~Neto}\ \emph {et~al.}(2009)\citenamefont
  {Castro~Neto}, \citenamefont {Guinea}, \citenamefont {Peres}, \citenamefont
  {Novoselov},\ and\ \citenamefont {Geim}}]{CastroNeto2009}%
  \BibitemOpen
  \bibfield  {author} {\bibinfo {author} {\bibfnamefont {A.~H.}\ \bibnamefont
  {Castro~Neto}}, \bibinfo {author} {\bibfnamefont {F.}~\bibnamefont {Guinea}},
  \bibinfo {author} {\bibfnamefont {N.~M.~R.}\ \bibnamefont {Peres}}, \bibinfo
  {author} {\bibfnamefont {K.~S.}\ \bibnamefont {Novoselov}}, \ and\ \bibinfo
  {author} {\bibfnamefont {A.~K.}\ \bibnamefont {Geim}},\ }\href {\doibase
  10.1103/RevModPhys.81.109} {\bibfield  {journal} {\bibinfo  {journal} {Rev.
  Mod. Phys.}\ }\textbf {\bibinfo {volume} {81}},\ \bibinfo {pages} {109}
  (\bibinfo {year} {2009})}\BibitemShut {NoStop}%
\bibitem [{\citenamefont {Yang}\ and\ \citenamefont
  {Nagaosa}(2014)}]{Yang2014}%
  \BibitemOpen
  \bibfield  {author} {\bibinfo {author} {\bibfnamefont {B.-J.}\ \bibnamefont
  {Yang}}\ and\ \bibinfo {author} {\bibfnamefont {N.}~\bibnamefont {Nagaosa}},\
  }\href {\doibase 10.1038/ncomms5898} {\bibfield  {journal} {\bibinfo
  {journal} {Nat. Commun.}\ }\textbf {\bibinfo {volume} {5}},\ \bibinfo {pages}
  {4898} (\bibinfo {year} {2014})}\BibitemShut {NoStop}%
\bibitem [{\citenamefont {Goh}\ and\ \citenamefont {Pickett}(2018)}]{Goh2018}%
  \BibitemOpen
  \bibfield  {author} {\bibinfo {author} {\bibfnamefont {W.~F.}\ \bibnamefont
  {Goh}}\ and\ \bibinfo {author} {\bibfnamefont {W.~E.}\ \bibnamefont
  {Pickett}},\ }\href {\doibase 10.1103/PhysRevB.97.035202} {\bibfield
  {journal} {\bibinfo  {journal} {Phys. Rev. B}\ }\textbf {\bibinfo {volume}
  {97}},\ \bibinfo {pages} {035202} (\bibinfo {year} {2018})}\BibitemShut
  {NoStop}%
\bibitem [{\citenamefont {Hsieh}, \citenamefont {Liu},\ and\ \citenamefont
  {Fu}(2014)}]{Hsieh2014}%
  \BibitemOpen
  \bibfield  {author} {\bibinfo {author} {\bibfnamefont {T.~H.}\ \bibnamefont
  {Hsieh}}, \bibinfo {author} {\bibfnamefont {J.}~\bibnamefont {Liu}}, \ and\
  \bibinfo {author} {\bibfnamefont {L.}~\bibnamefont {Fu}},\ }\href {\doibase
  10.1103/PhysRevB.90.081112} {\bibfield  {journal} {\bibinfo  {journal} {Phys.
  Rev. B}\ }\textbf {\bibinfo {volume} {90}},\ \bibinfo {pages} {081112}
  (\bibinfo {year} {2014})}\BibitemShut {NoStop}%
\bibitem [{\citenamefont {Suetsugu}\ \emph {et~al.}(2018)\citenamefont
  {Suetsugu}, \citenamefont {Hayama}, \citenamefont {Rost}, \citenamefont
  {Nuss}, \citenamefont {M{\"u}hle}, \citenamefont {Kim}, \citenamefont
  {Kitagawa},\ and\ \citenamefont {Takagi}}]{Suetsugu2018}%
  \BibitemOpen
  \bibfield  {author} {\bibinfo {author} {\bibfnamefont {S.}~\bibnamefont
  {Suetsugu}}, \bibinfo {author} {\bibfnamefont {K.}~\bibnamefont {Hayama}},
  \bibinfo {author} {\bibfnamefont {A.~W.}\ \bibnamefont {Rost}}, \bibinfo
  {author} {\bibfnamefont {J.}~\bibnamefont {Nuss}}, \bibinfo {author}
  {\bibfnamefont {C.}~\bibnamefont {M{\"u}hle}}, \bibinfo {author}
  {\bibfnamefont {J.}~\bibnamefont {Kim}}, \bibinfo {author} {\bibfnamefont
  {K.}~\bibnamefont {Kitagawa}}, \ and\ \bibinfo {author} {\bibfnamefont
  {H.}~\bibnamefont {Takagi}},\ }\href {\doibase 10.1103/PhysRevB.98.115203}
  {\bibfield  {journal} {\bibinfo  {journal} {Phys. Rev. B}\ }\textbf {\bibinfo
  {volume} {98}},\ \bibinfo {pages} {115203} (\bibinfo {year}
  {2018})}\BibitemShut {NoStop}%
\bibitem [{\citenamefont {Oudah}\ \emph {et~al.}(2016)\citenamefont {Oudah},
  \citenamefont {Ikeda}, \citenamefont {Hausmann}, \citenamefont {Yonezawa},
  \citenamefont {Fukumoto}, \citenamefont {Kobayashi}, \citenamefont {Sato},\
  and\ \citenamefont {Maeno}}]{Oudah2016}%
  \BibitemOpen
  \bibfield  {author} {\bibinfo {author} {\bibfnamefont {M.}~\bibnamefont
  {Oudah}}, \bibinfo {author} {\bibfnamefont {A.}~\bibnamefont {Ikeda}},
  \bibinfo {author} {\bibfnamefont {J.~N.}\ \bibnamefont {Hausmann}}, \bibinfo
  {author} {\bibfnamefont {S.}~\bibnamefont {Yonezawa}}, \bibinfo {author}
  {\bibfnamefont {T.}~\bibnamefont {Fukumoto}}, \bibinfo {author}
  {\bibfnamefont {S.}~\bibnamefont {Kobayashi}}, \bibinfo {author}
  {\bibfnamefont {M.}~\bibnamefont {Sato}}, \ and\ \bibinfo {author}
  {\bibfnamefont {Y.}~\bibnamefont {Maeno}},\ }\href {\doibase
  10.1038/ncomms13617} {\bibfield  {journal} {\bibinfo  {journal} {Nat.
  Commun.}\ }\textbf {\bibinfo {volume} {7}},\ \bibinfo {pages} {13617}
  (\bibinfo {year} {2016})}\BibitemShut {NoStop}%
\bibitem [{\citenamefont {Yu}\ \emph {et~al.}(2015)\citenamefont {Yu},
  \citenamefont {Weng}, \citenamefont {Fang}, \citenamefont {Dai},\ and\
  \citenamefont {Hu}}]{Yu2015}%
  \BibitemOpen
  \bibfield  {author} {\bibinfo {author} {\bibfnamefont {R.}~\bibnamefont
  {Yu}}, \bibinfo {author} {\bibfnamefont {H.}~\bibnamefont {Weng}}, \bibinfo
  {author} {\bibfnamefont {Z.}~\bibnamefont {Fang}}, \bibinfo {author}
  {\bibfnamefont {X.}~\bibnamefont {Dai}}, \ and\ \bibinfo {author}
  {\bibfnamefont {X.}~\bibnamefont {Hu}},\ }\href {\doibase
  10.1103/PhysRevLett.115.036807} {\bibfield  {journal} {\bibinfo  {journal}
  {Phys. Rev. Lett.}\ }\textbf {\bibinfo {volume} {115}},\ \bibinfo {pages}
  {036807} (\bibinfo {year} {2015})}\BibitemShut {NoStop}%
\bibitem [{\citenamefont {Chan}\ \emph {et~al.}(2016)\citenamefont {Chan},
  \citenamefont {Chiu}, \citenamefont {Chou},\ and\ \citenamefont
  {Schnyder}}]{Schnyder2016}%
  \BibitemOpen
  \bibfield  {author} {\bibinfo {author} {\bibfnamefont {Y.-H.}\ \bibnamefont
  {Chan}}, \bibinfo {author} {\bibfnamefont {C.-K.}\ \bibnamefont {Chiu}},
  \bibinfo {author} {\bibfnamefont {M.~Y.}\ \bibnamefont {Chou}}, \ and\
  \bibinfo {author} {\bibfnamefont {A.~P.}\ \bibnamefont {Schnyder}},\ }\href
  {\doibase 10.1103/PhysRevB.93.205132} {\bibfield  {journal} {\bibinfo
  {journal} {Phys. Rev. B}\ }\textbf {\bibinfo {volume} {93}},\ \bibinfo
  {pages} {205132} (\bibinfo {year} {2016})}\BibitemShut {NoStop}%
\bibitem [{\citenamefont {Butters}\ and\ \citenamefont
  {Myers}(1955)}]{Butters1955a}%
  \BibitemOpen
  \bibfield  {author} {\bibinfo {author} {\bibfnamefont {R.~G.}\ \bibnamefont
  {Butters}}\ and\ \bibinfo {author} {\bibfnamefont {H.~P.}\ \bibnamefont
  {Myers}},\ }\href {\doibase 10.1080/14786440208520557} {\bibfield  {journal}
  {\bibinfo  {journal} {Philos. Mag.}\ }\textbf {\bibinfo {volume} {46}},\
  \bibinfo {pages} {132} (\bibinfo {year} {1955})}\BibitemShut {NoStop}%
\bibitem [{\citenamefont {Brockhouse}\ and\ \citenamefont
  {Myers}(1957)}]{Brockhouse1957}%
  \BibitemOpen
  \bibfield  {author} {\bibinfo {author} {\bibfnamefont {B.~N.}\ \bibnamefont
  {Brockhouse}}\ and\ \bibinfo {author} {\bibfnamefont {H.~P.}\ \bibnamefont
  {Myers}},\ }\href {\doibase 10.1139/p57-035} {\bibfield  {journal} {\bibinfo
  {journal} {Can. J. Phys.}\ }\textbf {\bibinfo {volume} {35}},\ \bibinfo
  {pages} {313} (\bibinfo {year} {1957})}\BibitemShut {NoStop}%
\bibitem [{\citenamefont {Swanson}\ and\ \citenamefont
  {Friedberg}(1961)}]{Swanson1961}%
  \BibitemOpen
  \bibfield  {author} {\bibinfo {author} {\bibfnamefont {M.~L.}\ \bibnamefont
  {Swanson}}\ and\ \bibinfo {author} {\bibfnamefont {S.~A.}\ \bibnamefont
  {Friedberg}},\ }\href {\doibase 10.1139/p61-171} {\bibfield  {journal}
  {\bibinfo  {journal} {Can. J. Phys.}\ }\textbf {\bibinfo {volume} {39}},\
  \bibinfo {pages} {1429} (\bibinfo {year} {1961})}\BibitemShut {NoStop}%
\bibitem [{\citenamefont {Fruchart}\ and\ \citenamefont
  {Bertaut}(1978)}]{Fruchart1978}%
  \BibitemOpen
  \bibfield  {author} {\bibinfo {author} {\bibfnamefont {D.}~\bibnamefont
  {Fruchart}}\ and\ \bibinfo {author} {\bibfnamefont {F.}~\bibnamefont
  {Bertaut}},\ }\href {\doibase 10.1143/JPSJ.44.781} {\bibfield  {journal}
  {\bibinfo  {journal} {J. Phys. Soc. Jpn.}\ }\textbf {\bibinfo {volume}
  {44}},\ \bibinfo {pages} {781} (\bibinfo {year} {1978})}\BibitemShut
  {NoStop}%
\bibitem [{\citenamefont {Kaneko}, \citenamefont {Kanomata},\ and\
  \citenamefont {Shirakawa}(1987)}]{Kaneko1987a}%
  \BibitemOpen
  \bibfield  {author} {\bibinfo {author} {\bibfnamefont {T.}~\bibnamefont
  {Kaneko}}, \bibinfo {author} {\bibfnamefont {T.}~\bibnamefont {Kanomata}}, \
  and\ \bibinfo {author} {\bibfnamefont {K.}~\bibnamefont {Shirakawa}},\ }\href
  {\doibase 10.1143/JPSJ.56.4047} {\bibfield  {journal} {\bibinfo  {journal}
  {J. Phys. Soc. Jpn.}\ }\textbf {\bibinfo {volume} {56}},\ \bibinfo {pages}
  {4047} (\bibinfo {year} {1987})}\BibitemShut {NoStop}%
\bibitem [{\citenamefont {Fruchart}\ \emph {et~al.}(1973)\citenamefont
  {Fruchart}, \citenamefont {Bertaut}, \citenamefont {Clerc}, \citenamefont
  {Kh{\"o}i}, \citenamefont {Veillet}, \citenamefont {Lorthioir}, \citenamefont
  {Fruchart},\ and\ \citenamefont {Fruchart}}]{Fruchart1973}%
  \BibitemOpen
  \bibfield  {author} {\bibinfo {author} {\bibfnamefont {D.}~\bibnamefont
  {Fruchart}}, \bibinfo {author} {\bibfnamefont {E.}~\bibnamefont {Bertaut}},
  \bibinfo {author} {\bibfnamefont {B.~L.}\ \bibnamefont {Clerc}}, \bibinfo
  {author} {\bibfnamefont {L.~D.}\ \bibnamefont {Kh{\"o}i}}, \bibinfo {author}
  {\bibfnamefont {P.}~\bibnamefont {Veillet}}, \bibinfo {author} {\bibfnamefont
  {G.}~\bibnamefont {Lorthioir}}, \bibinfo {author} {\bibfnamefont
  {E.}~\bibnamefont {Fruchart}}, \ and\ \bibinfo {author} {\bibfnamefont
  {R.}~\bibnamefont {Fruchart}},\ }\href {\doibase
  10.1016/0022-4596(73)90012-1} {\bibfield  {journal} {\bibinfo  {journal} {J.
  Solid State Chem.}\ }\textbf {\bibinfo {volume} {8}},\ \bibinfo {pages} {182}
  (\bibinfo {year} {1973})}\BibitemShut {NoStop}%
\bibitem [{\citenamefont {Kaneko}\ \emph {et~al.}(1987)\citenamefont {Kaneko},
  \citenamefont {Kanomata}, \citenamefont {Miura}, \citenamefont {Kido},\ and\
  \citenamefont {Nakagawa}}]{Kaneko1987b}%
  \BibitemOpen
  \bibfield  {author} {\bibinfo {author} {\bibfnamefont {T.}~\bibnamefont
  {Kaneko}}, \bibinfo {author} {\bibfnamefont {T.}~\bibnamefont {Kanomata}},
  \bibinfo {author} {\bibfnamefont {S.}~\bibnamefont {Miura}}, \bibinfo
  {author} {\bibfnamefont {G.}~\bibnamefont {Kido}}, \ and\ \bibinfo {author}
  {\bibfnamefont {Y.}~\bibnamefont {Nakagawa}},\ }\href {\doibase
  10.1016/0304-8853(87)90431-8} {\bibfield  {journal} {\bibinfo  {journal} {J.
  Magn. Magn. Mater.}\ }\textbf {\bibinfo {volume} {70}},\ \bibinfo {pages}
  {261} (\bibinfo {year} {1987})}\BibitemShut {NoStop}%
\bibitem [{\citenamefont {Jardin}\ and\ \citenamefont
  {Labb{\'e}}(1975)}]{Jardin1975}%
  \BibitemOpen
  \bibfield  {author} {\bibinfo {author} {\bibfnamefont {J.~P.}\ \bibnamefont
  {Jardin}}\ and\ \bibinfo {author} {\bibfnamefont {J.}~\bibnamefont
  {Labb{\'e}}},\ }\href {\doibase 10.1051/jphys:0197500360120131700} {\bibfield
   {journal} {\bibinfo  {journal} {J. Phys. (France)}\ }\textbf {\bibinfo
  {volume} {36}},\ \bibinfo {pages} {1317} (\bibinfo {year}
  {1975})}\BibitemShut {NoStop}%
\bibitem [{\citenamefont {Motizuki}\ and\ \citenamefont
  {Nagai}(1988)}]{Motizuki1988}%
  \BibitemOpen
  \bibfield  {author} {\bibinfo {author} {\bibfnamefont {K.}~\bibnamefont
  {Motizuki}}\ and\ \bibinfo {author} {\bibfnamefont {H.}~\bibnamefont
  {Nagai}},\ }\href {\doibase 10.1088/0022-3719/21/30/011} {\bibfield
  {journal} {\bibinfo  {journal} {J. Phys. C}\ }\textbf {\bibinfo {volume}
  {21}},\ \bibinfo {pages} {5251} (\bibinfo {year} {1988})}\BibitemShut
  {NoStop}%
\bibitem [{\citenamefont {Antonov}\ \emph {et~al.}(2007)\citenamefont
  {Antonov}, \citenamefont {Harmon}, \citenamefont {Yaresko},\ and\
  \citenamefont {Shpak}}]{Antonov2007}%
  \BibitemOpen
  \bibfield  {author} {\bibinfo {author} {\bibfnamefont {V.~N.}\ \bibnamefont
  {Antonov}}, \bibinfo {author} {\bibfnamefont {B.~N.}\ \bibnamefont {Harmon}},
  \bibinfo {author} {\bibfnamefont {A.~N.}\ \bibnamefont {Yaresko}}, \ and\
  \bibinfo {author} {\bibfnamefont {A.~P.}\ \bibnamefont {Shpak}},\ }\href
  {\doibase 10.1103/PhysRevB.75.165114} {\bibfield  {journal} {\bibinfo
  {journal} {Phys. Rev. B}\ }\textbf {\bibinfo {volume} {75}},\ \bibinfo
  {pages} {165114} (\bibinfo {year} {2007})}\BibitemShut {NoStop}%
\bibitem [{\citenamefont {Gos{\'a}lbez-Mart{\'i}nez}, \citenamefont {Souza},\
  and\ \citenamefont {Vanderbilt}(2015)}]{GosalbezMartinez2015}%
  \BibitemOpen
  \bibfield  {author} {\bibinfo {author} {\bibfnamefont {D.}~\bibnamefont
  {Gos{\'a}lbez-Mart{\'i}nez}}, \bibinfo {author} {\bibfnamefont
  {I.}~\bibnamefont {Souza}}, \ and\ \bibinfo {author} {\bibfnamefont
  {D.}~\bibnamefont {Vanderbilt}},\ }\href {\doibase
  10.1103/PhysRevB.92.085138} {\bibfield  {journal} {\bibinfo  {journal} {Phys.
  Rev. B}\ }\textbf {\bibinfo {volume} {92}},\ \bibinfo {pages} {085138}
  (\bibinfo {year} {2015})}\BibitemShut {NoStop}%
\bibitem [{\citenamefont {Belopolski}\ \emph {et~al.}(2019)\citenamefont
  {Belopolski}, \citenamefont {Manna}, \citenamefont {Sanchez}, \citenamefont
  {Chang}, \citenamefont {Ernst}, \citenamefont {Yin}, \citenamefont {Zhang},
  \citenamefont {Cochran}, \citenamefont {Shumiya}, \citenamefont {Zheng},
  \citenamefont {Singh}, \citenamefont {Bian}, \citenamefont {Multer},
  \citenamefont {Litskevich}, \citenamefont {Zhou}, \citenamefont {Huang},
  \citenamefont {Wang}, \citenamefont {Chang}, \citenamefont {Xu},
  \citenamefont {Bansil}, \citenamefont {Felser}, \citenamefont {Lin},\ and\
  \citenamefont {Hasan}}]{Belopolski2019}%
  \BibitemOpen
  \bibfield  {author} {\bibinfo {author} {\bibfnamefont {I.}~\bibnamefont
  {Belopolski}}, \bibinfo {author} {\bibfnamefont {K.}~\bibnamefont {Manna}},
  \bibinfo {author} {\bibfnamefont {D.~S.}\ \bibnamefont {Sanchez}}, \bibinfo
  {author} {\bibfnamefont {G.}~\bibnamefont {Chang}}, \bibinfo {author}
  {\bibfnamefont {B.}~\bibnamefont {Ernst}}, \bibinfo {author} {\bibfnamefont
  {J.}~\bibnamefont {Yin}}, \bibinfo {author} {\bibfnamefont {S.~S.}\
  \bibnamefont {Zhang}}, \bibinfo {author} {\bibfnamefont {T.}~\bibnamefont
  {Cochran}}, \bibinfo {author} {\bibfnamefont {N.}~\bibnamefont {Shumiya}},
  \bibinfo {author} {\bibfnamefont {H.}~\bibnamefont {Zheng}}, \bibinfo
  {author} {\bibfnamefont {B.}~\bibnamefont {Singh}}, \bibinfo {author}
  {\bibfnamefont {G.}~\bibnamefont {Bian}}, \bibinfo {author} {\bibfnamefont
  {D.}~\bibnamefont {Multer}}, \bibinfo {author} {\bibfnamefont
  {M.}~\bibnamefont {Litskevich}}, \bibinfo {author} {\bibfnamefont
  {X.}~\bibnamefont {Zhou}}, \bibinfo {author} {\bibfnamefont {S.-M.}\
  \bibnamefont {Huang}}, \bibinfo {author} {\bibfnamefont {B.}~\bibnamefont
  {Wang}}, \bibinfo {author} {\bibfnamefont {T.-R.}\ \bibnamefont {Chang}},
  \bibinfo {author} {\bibfnamefont {S.-Y.}\ \bibnamefont {Xu}}, \bibinfo
  {author} {\bibfnamefont {A.}~\bibnamefont {Bansil}}, \bibinfo {author}
  {\bibfnamefont {C.}~\bibnamefont {Felser}}, \bibinfo {author} {\bibfnamefont
  {H.}~\bibnamefont {Lin}}, \ and\ \bibinfo {author} {\bibfnamefont {M.~Z.}\
  \bibnamefont {Hasan}},\ }\href {\doibase 10.1126/science.aav2327} {\bibfield
  {journal} {\bibinfo  {journal} {Science}\ }\textbf {\bibinfo {volume}
  {365}},\ \bibinfo {pages} {1278} (\bibinfo {year} {2019})}\BibitemShut
  {NoStop}%
\bibitem [{\citenamefont {Kresse}\ and\ \citenamefont
  {Hafner}(1994)}]{Kresse1994}%
  \BibitemOpen
  \bibfield  {author} {\bibinfo {author} {\bibfnamefont {G.}~\bibnamefont
  {Kresse}}\ and\ \bibinfo {author} {\bibfnamefont {J.}~\bibnamefont
  {Hafner}},\ }\href {\doibase 10.1103/PhysRevB.49.14251} {\bibfield  {journal}
  {\bibinfo  {journal} {Phys. Rev. B}\ }\textbf {\bibinfo {volume} {49}},\
  \bibinfo {pages} {14251} (\bibinfo {year} {1994})}\BibitemShut {NoStop}%
\bibitem [{\citenamefont {Kresse}\ and\ \citenamefont
  {Furthm{\"u}ller}(1996{\natexlab{a}})}]{Kresse1996a}%
  \BibitemOpen
  \bibfield  {author} {\bibinfo {author} {\bibfnamefont {G.}~\bibnamefont
  {Kresse}}\ and\ \bibinfo {author} {\bibfnamefont {J.}~\bibnamefont
  {Furthm{\"u}ller}},\ }\href {\doibase 10.1103/PhysRevB.54.11169} {\bibfield
  {journal} {\bibinfo  {journal} {Phys. Rev. B}\ }\textbf {\bibinfo {volume}
  {54}},\ \bibinfo {pages} {11169} (\bibinfo {year}
  {1996}{\natexlab{a}})}\BibitemShut {NoStop}%
\bibitem [{\citenamefont {Kresse}\ and\ \citenamefont
  {Furthm{\"u}ller}(1996{\natexlab{b}})}]{Kresse1996b}%
  \BibitemOpen
  \bibfield  {author} {\bibinfo {author} {\bibfnamefont {G.}~\bibnamefont
  {Kresse}}\ and\ \bibinfo {author} {\bibfnamefont {J.}~\bibnamefont
  {Furthm{\"u}ller}},\ }\href {\doibase 10.1016/0927-0256(96)00008-0}
  {\bibfield  {journal} {\bibinfo  {journal} {Comput. Mater. Sci.}\ }\textbf
  {\bibinfo {volume} {6}},\ \bibinfo {pages} {15} (\bibinfo {year}
  {1996}{\natexlab{b}})}\BibitemShut {NoStop}%
\bibitem [{\citenamefont {Blaha}\ \emph {et~al.}(2001)\citenamefont {Blaha},
  \citenamefont {Schwarz}, \citenamefont {Madsen}, \citenamefont {Kvasnicka},\
  and\ \citenamefont {Luitz}}]{Blaha2001}%
  \BibitemOpen
  \bibfield  {author} {\bibinfo {author} {\bibfnamefont {P.}~\bibnamefont
  {Blaha}}, \bibinfo {author} {\bibfnamefont {K.}~\bibnamefont {Schwarz}},
  \bibinfo {author} {\bibfnamefont {G.}~\bibnamefont {Madsen}}, \bibinfo
  {author} {\bibfnamefont {D.}~\bibnamefont {Kvasnicka}}, \ and\ \bibinfo
  {author} {\bibfnamefont {J.}~\bibnamefont {Luitz}},\ }\href
  {http://susi.theochem.tuwien.ac.at/reg_user/textbooks/usersguide.pdf} {\emph
  {\bibinfo {title} {{WIEN2K}: An Augmented Plane Wave and Local Orbitals
  Program for Calculating Crystal Properties}}}\ (\bibinfo  {publisher}
  {Karlheinz Schwarz},\ \bibinfo {year} {2001})\BibitemShut {NoStop}%
\bibitem [{\citenamefont {Perdew}, \citenamefont {Burke},\ and\ \citenamefont
  {Ernzerhof}(1996)}]{Perdew1996}%
  \BibitemOpen
  \bibfield  {author} {\bibinfo {author} {\bibfnamefont {J.~P.}\ \bibnamefont
  {Perdew}}, \bibinfo {author} {\bibfnamefont {K.}~\bibnamefont {Burke}}, \
  and\ \bibinfo {author} {\bibfnamefont {M.}~\bibnamefont {Ernzerhof}},\ }\href
  {\doibase 10.1103/PhysRevLett.77.3865} {\bibfield  {journal} {\bibinfo
  {journal} {Phys. Rev. Lett.}\ }\textbf {\bibinfo {volume} {77}},\ \bibinfo
  {pages} {3865} (\bibinfo {year} {1996})}\BibitemShut {NoStop}%
\bibitem [{\citenamefont {Bl{\"o}chl}(1994)}]{Blochl1994a}%
  \BibitemOpen
  \bibfield  {author} {\bibinfo {author} {\bibfnamefont {P.~E.}\ \bibnamefont
  {Bl{\"o}chl}},\ }\href {\doibase 10.1103/PhysRevB.50.17953} {\bibfield
  {journal} {\bibinfo  {journal} {Phys. Rev. B}\ }\textbf {\bibinfo {volume}
  {50}},\ \bibinfo {pages} {17953} (\bibinfo {year} {1994})}\BibitemShut
  {NoStop}%
\bibitem [{\citenamefont {Kresse}\ and\ \citenamefont
  {Joubert}(1999)}]{Kresse1999}%
  \BibitemOpen
  \bibfield  {author} {\bibinfo {author} {\bibfnamefont {G.}~\bibnamefont
  {Kresse}}\ and\ \bibinfo {author} {\bibfnamefont {D.}~\bibnamefont
  {Joubert}},\ }\href {\doibase 10.1103/PhysRevB.59.1758} {\bibfield  {journal}
  {\bibinfo  {journal} {Phys. Rev. B}\ }\textbf {\bibinfo {volume} {59}},\
  \bibinfo {pages} {1758} (\bibinfo {year} {1999})}\BibitemShut {NoStop}%
\bibitem [{\citenamefont {Schwarz}, \citenamefont {Blaha},\ and\ \citenamefont
  {Madsen}(2002)}]{Schwarz2002}%
  \BibitemOpen
  \bibfield  {author} {\bibinfo {author} {\bibfnamefont {K.}~\bibnamefont
  {Schwarz}}, \bibinfo {author} {\bibfnamefont {P.}~\bibnamefont {Blaha}}, \
  and\ \bibinfo {author} {\bibfnamefont {G.}~\bibnamefont {Madsen}},\ }\href
  {\doibase 10.1016/S0010-4655(02)00206-0} {\bibfield  {journal} {\bibinfo
  {journal} {Comput. Phys. Commun.}\ }\textbf {\bibinfo {volume} {147}},\
  \bibinfo {pages} {71} (\bibinfo {year} {2002})}\BibitemShut {NoStop}%
\bibitem [{\citenamefont {Bl{\"o}chl}, \citenamefont {Jepsen},\ and\
  \citenamefont {Andersen}(1994)}]{Blochl1994b}%
  \BibitemOpen
  \bibfield  {author} {\bibinfo {author} {\bibfnamefont {P.~E.}\ \bibnamefont
  {Bl{\"o}chl}}, \bibinfo {author} {\bibfnamefont {O.}~\bibnamefont {Jepsen}},
  \ and\ \bibinfo {author} {\bibfnamefont {O.~K.}\ \bibnamefont {Andersen}},\
  }\href {\doibase 10.1103/PhysRevB.49.16223} {\bibfield  {journal} {\bibinfo
  {journal} {Phys. Rev. B}\ }\textbf {\bibinfo {volume} {49}},\ \bibinfo
  {pages} {16223} (\bibinfo {year} {1994})}\BibitemShut {NoStop}%
\bibitem [{\citenamefont {Mostofi}\ \emph {et~al.}(2014)\citenamefont
  {Mostofi}, \citenamefont {Yates}, \citenamefont {Pizzi}, \citenamefont {Lee},
  \citenamefont {Souza}, \citenamefont {Vanderbilt},\ and\ \citenamefont
  {Marzari}}]{Mostofi2014}%
  \BibitemOpen
  \bibfield  {author} {\bibinfo {author} {\bibfnamefont {A.~A.}\ \bibnamefont
  {Mostofi}}, \bibinfo {author} {\bibfnamefont {J.~R.}\ \bibnamefont {Yates}},
  \bibinfo {author} {\bibfnamefont {G.}~\bibnamefont {Pizzi}}, \bibinfo
  {author} {\bibfnamefont {Y.-S.}\ \bibnamefont {Lee}}, \bibinfo {author}
  {\bibfnamefont {I.}~\bibnamefont {Souza}}, \bibinfo {author} {\bibfnamefont
  {D.}~\bibnamefont {Vanderbilt}}, \ and\ \bibinfo {author} {\bibfnamefont
  {N.}~\bibnamefont {Marzari}},\ }\href {\doibase 10.1016/j.cpc.2014.05.003}
  {\bibfield  {journal} {\bibinfo  {journal} {Comput. Phys. Commun.}\ }\textbf
  {\bibinfo {volume} {185}},\ \bibinfo {pages} {2309} (\bibinfo {year}
  {2014})}\BibitemShut {NoStop}%
\bibitem [{\citenamefont {Wu}\ \emph {et~al.}(2018)\citenamefont {Wu},
  \citenamefont {Zhang}, \citenamefont {Song}, \citenamefont {Troyer},\ and\
  \citenamefont {Soluyanov}}]{Wu2018}%
  \BibitemOpen
  \bibfield  {author} {\bibinfo {author} {\bibfnamefont {Q.-S.}\ \bibnamefont
  {Wu}}, \bibinfo {author} {\bibfnamefont {S.-N.}\ \bibnamefont {Zhang}},
  \bibinfo {author} {\bibfnamefont {H.-F.}\ \bibnamefont {Song}}, \bibinfo
  {author} {\bibfnamefont {M.}~\bibnamefont {Troyer}}, \ and\ \bibinfo {author}
  {\bibfnamefont {A.~A.}\ \bibnamefont {Soluyanov}},\ }\href {\doibase
  10.1016/j.cpc.2017.09.033} {\bibfield  {journal} {\bibinfo  {journal}
  {Comput. Phys. Commun.}\ }\textbf {\bibinfo {volume} {224}},\ \bibinfo
  {pages} {405} (\bibinfo {year} {2018})}\BibitemShut {NoStop}%
\bibitem [{\citenamefont {Sancho}\ \emph {et~al.}(1985)\citenamefont {Sancho},
  \citenamefont {Sancho}, \citenamefont {Sancho},\ and\ \citenamefont
  {Rubio}}]{Sancho1985}%
  \BibitemOpen
  \bibfield  {author} {\bibinfo {author} {\bibfnamefont {M.~P.~L.}\
  \bibnamefont {Sancho}}, \bibinfo {author} {\bibfnamefont {J.~M.~L.}\
  \bibnamefont {Sancho}}, \bibinfo {author} {\bibfnamefont {J.~M.~L.}\
  \bibnamefont {Sancho}}, \ and\ \bibinfo {author} {\bibfnamefont
  {J.}~\bibnamefont {Rubio}},\ }\href {\doibase 10.1088/0305-4608/15/4/009}
  {\bibfield  {journal} {\bibinfo  {journal} {J. Phys. F}\ }\textbf {\bibinfo
  {volume} {15}},\ \bibinfo {pages} {851} (\bibinfo {year} {1985})}\BibitemShut
  {NoStop}%
\bibitem [{\citenamefont {Medeiros}, \citenamefont {Stafstr{\"o}m},\ and\
  \citenamefont {Bj{\"o}rk}(2014)}]{Medeiros2014}%
  \BibitemOpen
  \bibfield  {author} {\bibinfo {author} {\bibfnamefont {P.~V.~C.}\
  \bibnamefont {Medeiros}}, \bibinfo {author} {\bibfnamefont {S.}~\bibnamefont
  {Stafstr{\"o}m}}, \ and\ \bibinfo {author} {\bibfnamefont {J.}~\bibnamefont
  {Bj{\"o}rk}},\ }\href {\doibase 10.1103/PhysRevB.89.041407} {\bibfield
  {journal} {\bibinfo  {journal} {Phys. Rev. B}\ }\textbf {\bibinfo {volume}
  {89}},\ \bibinfo {pages} {041407} (\bibinfo {year} {2014})}\BibitemShut
  {NoStop}%
\bibitem [{\citenamefont {Medeiros}\ \emph {et~al.}(2015)\citenamefont
  {Medeiros}, \citenamefont {Tsirkin}, \citenamefont {Stafstr{\"o}m},\ and\
  \citenamefont {Bj{\"o}rk}}]{Medeiros2015}%
  \BibitemOpen
  \bibfield  {author} {\bibinfo {author} {\bibfnamefont {P.~V.~C.}\
  \bibnamefont {Medeiros}}, \bibinfo {author} {\bibfnamefont {S.~S.}\
  \bibnamefont {Tsirkin}}, \bibinfo {author} {\bibfnamefont {S.}~\bibnamefont
  {Stafstr{\"o}m}}, \ and\ \bibinfo {author} {\bibfnamefont {J.}~\bibnamefont
  {Bj{\"o}rk}},\ }\href {\doibase 10.1103/PhysRevB.91.041116} {\bibfield
  {journal} {\bibinfo  {journal} {Phys. Rev. B}\ }\textbf {\bibinfo {volume}
  {91}},\ \bibinfo {pages} {041116} (\bibinfo {year} {2015})}\BibitemShut
  {NoStop}%
\bibitem [{\citenamefont {Dronskowski}\ and\ \citenamefont
  {Bloechl}(1993)}]{Dronskowski1993}%
  \BibitemOpen
  \bibfield  {author} {\bibinfo {author} {\bibfnamefont {R.}~\bibnamefont
  {Dronskowski}}\ and\ \bibinfo {author} {\bibfnamefont {P.~E.}\ \bibnamefont
  {Bloechl}},\ }\href {\doibase 10.1021/j100135a014} {\bibfield  {journal}
  {\bibinfo  {journal} {J. Phys. Chem.}\ }\textbf {\bibinfo {volume} {97}},\
  \bibinfo {pages} {8617} (\bibinfo {year} {1993})}\BibitemShut {NoStop}%
\bibitem [{\citenamefont {Deringer}, \citenamefont {Tchougr{\'e}eff},\ and\
  \citenamefont {Dronskowski}(2011)}]{Deringer2011}%
  \BibitemOpen
  \bibfield  {author} {\bibinfo {author} {\bibfnamefont {V.~L.}\ \bibnamefont
  {Deringer}}, \bibinfo {author} {\bibfnamefont {A.~L.}\ \bibnamefont
  {Tchougr{\'e}eff}}, \ and\ \bibinfo {author} {\bibfnamefont {R.}~\bibnamefont
  {Dronskowski}},\ }\href {\doibase 10.1021/jp202489s} {\bibfield  {journal}
  {\bibinfo  {journal} {J. Phys. Chem. A}\ }\textbf {\bibinfo {volume} {115}},\
  \bibinfo {pages} {5461} (\bibinfo {year} {2011})}\BibitemShut {NoStop}%
\bibitem [{\citenamefont {Maintz}\ \emph {et~al.}(2013)\citenamefont {Maintz},
  \citenamefont {Deringer}, \citenamefont {Tchougr{\'e}eff},\ and\
  \citenamefont {Dronskowski}}]{Maintz2013}%
  \BibitemOpen
  \bibfield  {author} {\bibinfo {author} {\bibfnamefont {S.}~\bibnamefont
  {Maintz}}, \bibinfo {author} {\bibfnamefont {V.~L.}\ \bibnamefont
  {Deringer}}, \bibinfo {author} {\bibfnamefont {A.~L.}\ \bibnamefont
  {Tchougr{\'e}eff}}, \ and\ \bibinfo {author} {\bibfnamefont {R.}~\bibnamefont
  {Dronskowski}},\ }\href {\doibase 10.1002/jcc.23424} {\bibfield  {journal}
  {\bibinfo  {journal} {J. Comput. Chem.}\ }\textbf {\bibinfo {volume} {34}},\
  \bibinfo {pages} {2557} (\bibinfo {year} {2013})}\BibitemShut {NoStop}%
\bibitem [{\citenamefont {Maintz}\ \emph {et~al.}(2016)\citenamefont {Maintz},
  \citenamefont {Deringer}, \citenamefont {Tchougr{\'e}eff},\ and\
  \citenamefont {Dronskowski}}]{Maintz2016}%
  \BibitemOpen
  \bibfield  {author} {\bibinfo {author} {\bibfnamefont {S.}~\bibnamefont
  {Maintz}}, \bibinfo {author} {\bibfnamefont {V.~L.}\ \bibnamefont
  {Deringer}}, \bibinfo {author} {\bibfnamefont {A.~L.}\ \bibnamefont
  {Tchougr{\'e}eff}}, \ and\ \bibinfo {author} {\bibfnamefont {R.}~\bibnamefont
  {Dronskowski}},\ }\href {\doibase 10.1002/jcc.24300} {\bibfield  {journal}
  {\bibinfo  {journal} {J. Comput. Chem.}\ }\textbf {\bibinfo {volume} {37}},\
  \bibinfo {pages} {1030} (\bibinfo {year} {2016})}\BibitemShut {NoStop}%
\bibitem [{\citenamefont {Momma}\ and\ \citenamefont
  {Izumi}(2011)}]{Momma2011}%
  \BibitemOpen
  \bibfield  {author} {\bibinfo {author} {\bibfnamefont {K.}~\bibnamefont
  {Momma}}\ and\ \bibinfo {author} {\bibfnamefont {F.}~\bibnamefont {Izumi}},\
  }\href {\doibase 10.1107/S0021889811038970} {\bibfield  {journal} {\bibinfo
  {journal} {J. Appl. Crystallogr.}\ }\textbf {\bibinfo {volume} {44}},\
  \bibinfo {pages} {1272} (\bibinfo {year} {2011})}\BibitemShut {NoStop}%
\bibitem [{\citenamefont {Bocarsly}\ \emph {et~al.}(2018)\citenamefont
  {Bocarsly}, \citenamefont {Need}, \citenamefont {Seshadri},\ and\
  \citenamefont {Wilson}}]{Bocarsly2018}%
  \BibitemOpen
  \bibfield  {author} {\bibinfo {author} {\bibfnamefont {J.~D.}\ \bibnamefont
  {Bocarsly}}, \bibinfo {author} {\bibfnamefont {R.~F.}\ \bibnamefont {Need}},
  \bibinfo {author} {\bibfnamefont {R.}~\bibnamefont {Seshadri}}, \ and\
  \bibinfo {author} {\bibfnamefont {S.~D.}\ \bibnamefont {Wilson}},\ }\href
  {\doibase 10.1103/PhysRevB.97.100404} {\bibfield  {journal} {\bibinfo
  {journal} {Phys. Rev. B}\ }\textbf {\bibinfo {volume} {97}},\ \bibinfo
  {pages} {100404} (\bibinfo {year} {2018})}\BibitemShut {NoStop}%
\bibitem [{\citenamefont {Dronskowski}\ \emph {et~al.}(2002)\citenamefont
  {Dronskowski}, \citenamefont {Korczak}, \citenamefont {Lueken},\ and\
  \citenamefont {Jung}}]{Dronskowski2002}%
  \BibitemOpen
  \bibfield  {author} {\bibinfo {author} {\bibfnamefont {R.}~\bibnamefont
  {Dronskowski}}, \bibinfo {author} {\bibfnamefont {K.}~\bibnamefont
  {Korczak}}, \bibinfo {author} {\bibfnamefont {H.}~\bibnamefont {Lueken}}, \
  and\ \bibinfo {author} {\bibfnamefont {W.}~\bibnamefont {Jung}},\ }\href
  {\doibase 10.1002/1521-3773(20020715)41:14<2528::AID-ANIE2528>3.0.CO;2-6}
  {\bibfield  {journal} {\bibinfo  {journal} {Angew. Chem. Int. Ed.}\ }\textbf
  {\bibinfo {volume} {41}},\ \bibinfo {pages} {2528} (\bibinfo {year}
  {2002})}\BibitemShut {NoStop}%
\bibitem [{\citenamefont {Hoffmann}(1987)}]{Hoffmann1987}%
  \BibitemOpen
  \bibfield  {author} {\bibinfo {author} {\bibfnamefont {R.}~\bibnamefont
  {Hoffmann}},\ }\href {\doibase 10.1002/anie.198708461} {\bibfield  {journal}
  {\bibinfo  {journal} {Angew. Chem. Int. Ed.}\ }\textbf {\bibinfo {volume}
  {26}},\ \bibinfo {pages} {846} (\bibinfo {year} {1987})}\BibitemShut
  {NoStop}%
\bibitem [{\citenamefont {Wang}\ \emph {et~al.}(2018)\citenamefont {Wang},
  \citenamefont {Liu}, \citenamefont {Jin}, \citenamefont {Sui}, \citenamefont
  {Zhang}, \citenamefont {Duan}, \citenamefont {Liu},\ and\ \citenamefont
  {Huang}}]{Wang2018}%
  \BibitemOpen
  \bibfield  {author} {\bibinfo {author} {\bibfnamefont {J.}~\bibnamefont
  {Wang}}, \bibinfo {author} {\bibfnamefont {Y.}~\bibnamefont {Liu}}, \bibinfo
  {author} {\bibfnamefont {K.-H.}\ \bibnamefont {Jin}}, \bibinfo {author}
  {\bibfnamefont {X.}~\bibnamefont {Sui}}, \bibinfo {author} {\bibfnamefont
  {L.}~\bibnamefont {Zhang}}, \bibinfo {author} {\bibfnamefont
  {W.}~\bibnamefont {Duan}}, \bibinfo {author} {\bibfnamefont {F.}~\bibnamefont
  {Liu}}, \ and\ \bibinfo {author} {\bibfnamefont {B.}~\bibnamefont {Huang}},\
  }\href {\doibase 10.1103/PhysRevB.98.201112} {\bibfield  {journal} {\bibinfo
  {journal} {Phys. Rev. B}\ }\textbf {\bibinfo {volume} {98}},\ \bibinfo
  {pages} {201112} (\bibinfo {year} {2018})}\BibitemShut {NoStop}%
\bibitem [{\citenamefont {Teicher}\ \emph {et~al.}(2019)\citenamefont
  {Teicher}, \citenamefont {Lamontagne}, \citenamefont {Schoop},\ and\
  \citenamefont {Seshadri}}]{Teicher2019a}%
  \BibitemOpen
  \bibfield  {author} {\bibinfo {author} {\bibfnamefont {S.~M.~L.}\
  \bibnamefont {Teicher}}, \bibinfo {author} {\bibfnamefont {L.~K.}\
  \bibnamefont {Lamontagne}}, \bibinfo {author} {\bibfnamefont {L.~M.}\
  \bibnamefont {Schoop}}, \ and\ \bibinfo {author} {\bibfnamefont
  {R.}~\bibnamefont {Seshadri}},\ }\href {\doibase 10.1103/PhysRevB.99.195148}
  {\bibfield  {journal} {\bibinfo  {journal} {Phys. Rev. B}\ }\textbf {\bibinfo
  {volume} {99}},\ \bibinfo {pages} {195148} (\bibinfo {year}
  {2019})}\BibitemShut {NoStop}%
\bibitem [{\citenamefont {Wang}\ \emph {et~al.}(2016)\citenamefont {Wang},
  \citenamefont {Vergniory}, \citenamefont {Kushwaha}, \citenamefont
  {Hirschberger}, \citenamefont {Chulkov}, \citenamefont {Ernst}, \citenamefont
  {Ong}, \citenamefont {Cava},\ and\ \citenamefont {Bernevig}}]{Wang2016}%
  \BibitemOpen
  \bibfield  {author} {\bibinfo {author} {\bibfnamefont {Z.}~\bibnamefont
  {Wang}}, \bibinfo {author} {\bibfnamefont {M.~G.}\ \bibnamefont {Vergniory}},
  \bibinfo {author} {\bibfnamefont {S.}~\bibnamefont {Kushwaha}}, \bibinfo
  {author} {\bibfnamefont {M.}~\bibnamefont {Hirschberger}}, \bibinfo {author}
  {\bibfnamefont {E.~V.}\ \bibnamefont {Chulkov}}, \bibinfo {author}
  {\bibfnamefont {A.}~\bibnamefont {Ernst}}, \bibinfo {author} {\bibfnamefont
  {N.~P.}\ \bibnamefont {Ong}}, \bibinfo {author} {\bibfnamefont {R.~J.}\
  \bibnamefont {Cava}}, \ and\ \bibinfo {author} {\bibfnamefont {B.~A.}\
  \bibnamefont {Bernevig}},\ }\href {\doibase 10.1103/PhysRevLett.117.236401}
  {\bibfield  {journal} {\bibinfo  {journal} {Phys. Rev. Lett.}\ }\textbf
  {\bibinfo {volume} {117}},\ \bibinfo {pages} {236401} (\bibinfo {year}
  {2016})}\BibitemShut {NoStop}%
\bibitem [{\citenamefont {Vanderbilt}(2018)}]{Vanderbilt2018}%
  \BibitemOpen
  \bibfield  {author} {\bibinfo {author} {\bibfnamefont {D.}~\bibnamefont
  {Vanderbilt}},\ }\href@noop {} {\emph {\bibinfo {title} {Berry Phases in
  Electronic Structure Theory}}}\ (\bibinfo  {publisher} {Cambridge University
  Press},\ \bibinfo {year} {2018})\BibitemShut {NoStop}%
\bibitem [{\citenamefont {Armitage}, \citenamefont {Mele},\ and\ \citenamefont
  {Vishwanath}(2018)}]{Armitage2018}%
  \BibitemOpen
  \bibfield  {author} {\bibinfo {author} {\bibfnamefont {N.~P.}\ \bibnamefont
  {Armitage}}, \bibinfo {author} {\bibfnamefont {E.~J.}\ \bibnamefont {Mele}},
  \ and\ \bibinfo {author} {\bibfnamefont {A.}~\bibnamefont {Vishwanath}},\
  }\href {\doibase 10.1103/RevModPhys.90.015001} {\bibfield  {journal}
  {\bibinfo  {journal} {Rev. Mod. Phys.}\ }\textbf {\bibinfo {volume} {90}},\
  \bibinfo {pages} {015001} (\bibinfo {year} {2018})}\BibitemShut {NoStop}%
\bibitem [{\citenamefont {Karen}\ \emph {et~al.}(1991)\citenamefont {Karen},
  \citenamefont {Fjellv{\aa}g}, \citenamefont {Kjekshus},\ and\ \citenamefont
  {Andresen}}]{Karen1991}%
  \BibitemOpen
  \bibfield  {author} {\bibinfo {author} {\bibfnamefont {P.}~\bibnamefont
  {Karen}}, \bibinfo {author} {\bibfnamefont {H.}~\bibnamefont {Fjellv{\aa}g}},
  \bibinfo {author} {\bibfnamefont {A.}~\bibnamefont {Kjekshus}}, \ and\
  \bibinfo {author} {\bibfnamefont {A.~F.}\ \bibnamefont {Andresen}},\ }\href
  {\doibase 10.3891/acta.chem.scand.45-0549} {\bibfield  {journal} {\bibinfo
  {journal} {Acta Chem. Scand.}\ }\textbf {\bibinfo {volume} {45}},\ \bibinfo
  {pages} {549} (\bibinfo {year} {1991})}\BibitemShut {NoStop}%
\bibitem [{\citenamefont {Franco}\ \emph {et~al.}(2012)\citenamefont {Franco},
  \citenamefont {Bl{\'a}zquez}, \citenamefont {Ingale},\ and\ \citenamefont
  {Conde}}]{Franco2012}%
  \BibitemOpen
  \bibfield  {author} {\bibinfo {author} {\bibfnamefont {V.}~\bibnamefont
  {Franco}}, \bibinfo {author} {\bibfnamefont {J.}~\bibnamefont
  {Bl{\'a}zquez}}, \bibinfo {author} {\bibfnamefont {B.}~\bibnamefont
  {Ingale}}, \ and\ \bibinfo {author} {\bibfnamefont {A.}~\bibnamefont
  {Conde}},\ }\href {\doibase 10.1146/annurev-matsci-062910-100356} {\bibfield
  {journal} {\bibinfo  {journal} {Annu. Rev. Mater. Res.}\ }\textbf {\bibinfo
  {volume} {42}},\ \bibinfo {pages} {305} (\bibinfo {year} {2012})}\BibitemShut
  {NoStop}%
\bibitem [{\citenamefont {Park}\ \emph {et~al.}(2019)\citenamefont {Park},
  \citenamefont {Lee}, \citenamefont {Kim}, \citenamefont {Jin},\ and\
  \citenamefont {Yang}}]{Park2019}%
  \BibitemOpen
  \bibfield  {author} {\bibinfo {author} {\bibfnamefont {J.-H.}\ \bibnamefont
  {Park}}, \bibinfo {author} {\bibfnamefont {S.~H.}\ \bibnamefont {Lee}},
  \bibinfo {author} {\bibfnamefont {C.~H.}\ \bibnamefont {Kim}}, \bibinfo
  {author} {\bibfnamefont {H.}~\bibnamefont {Jin}}, \ and\ \bibinfo {author}
  {\bibfnamefont {B.-J.}\ \bibnamefont {Yang}},\ }\href {\doibase
  10.1103/PhysRevB.99.195107} {\bibfield  {journal} {\bibinfo  {journal} {Phys.
  Rev. B}\ }\textbf {\bibinfo {volume} {99}},\ \bibinfo {pages} {195107}
  (\bibinfo {year} {2019})}\BibitemShut {NoStop}%
\bibitem [{\citenamefont {Decker}, \citenamefont {Landrum},\ and\ \citenamefont
  {Dronskowski}(2002)}]{Decker2002}%
  \BibitemOpen
  \bibfield  {author} {\bibinfo {author} {\bibfnamefont {A.}~\bibnamefont
  {Decker}}, \bibinfo {author} {\bibfnamefont {G.~A.}\ \bibnamefont {Landrum}},
  \ and\ \bibinfo {author} {\bibfnamefont {R.}~\bibnamefont {Dronskowski}},\
  }\href {\doibase 10.1002/1521-3749(200201)628:1<303::AID-ZAAC303>3.0.CO;2-W}
  {\bibfield  {journal} {\bibinfo  {journal} {Z. Anorg. Allg. Chem.}\ }\textbf
  {\bibinfo {volume} {628}},\ \bibinfo {pages} {303} (\bibinfo {year}
  {2002})}\BibitemShut {NoStop}%
\bibitem [{\citenamefont {Johannes}\ and\ \citenamefont
  {Mazin}(2008)}]{Johannes2008}%
  \BibitemOpen
  \bibfield  {author} {\bibinfo {author} {\bibfnamefont {M.~D.}\ \bibnamefont
  {Johannes}}\ and\ \bibinfo {author} {\bibfnamefont {I.~I.}\ \bibnamefont
  {Mazin}},\ }\href {\doibase 10.1103/PhysRevB.77.165135} {\bibfield  {journal}
  {\bibinfo  {journal} {Phys. Rev. B}\ }\textbf {\bibinfo {volume} {77}},\
  \bibinfo {pages} {165135} (\bibinfo {year} {2008})}\BibitemShut {NoStop}%
\bibitem [{\citenamefont {{Shi}}\ \emph {et~al.}(2019)\citenamefont {{Shi}},
  \citenamefont {{Wieder}}, \citenamefont {{Meyerheim}}, \citenamefont {{Sun}},
  \citenamefont {{Zhang}}, \citenamefont {{Li}}, \citenamefont {{Shen}},
  \citenamefont {{Qi}}, \citenamefont {{Yang}}, \citenamefont {{Jena}},
  \citenamefont {{Werner}}, \citenamefont {{Koepernik}}, \citenamefont
  {{Parkin}}, \citenamefont {{Chen}}, \citenamefont {{Felser}}, \citenamefont
  {{Bernevig}},\ and\ \citenamefont {{Wang}}}]{Shi2018}%
  \BibitemOpen
  \bibfield  {author} {\bibinfo {author} {\bibfnamefont {W.}~\bibnamefont
  {{Shi}}}, \bibinfo {author} {\bibfnamefont {B.~J.}\ \bibnamefont {{Wieder}}},
  \bibinfo {author} {\bibfnamefont {H.~L.}\ \bibnamefont {{Meyerheim}}},
  \bibinfo {author} {\bibfnamefont {Y.}~\bibnamefont {{Sun}}}, \bibinfo
  {author} {\bibfnamefont {Y.}~\bibnamefont {{Zhang}}}, \bibinfo {author}
  {\bibfnamefont {Y.}~\bibnamefont {{Li}}}, \bibinfo {author} {\bibfnamefont
  {L.}~\bibnamefont {{Shen}}}, \bibinfo {author} {\bibfnamefont
  {Y.}~\bibnamefont {{Qi}}}, \bibinfo {author} {\bibfnamefont {L.}~\bibnamefont
  {{Yang}}}, \bibinfo {author} {\bibfnamefont {J.}~\bibnamefont {{Jena}}},
  \bibinfo {author} {\bibfnamefont {P.}~\bibnamefont {{Werner}}}, \bibinfo
  {author} {\bibfnamefont {K.}~\bibnamefont {{Koepernik}}}, \bibinfo {author}
  {\bibfnamefont {S.}~\bibnamefont {{Parkin}}}, \bibinfo {author}
  {\bibfnamefont {Y.}~\bibnamefont {{Chen}}}, \bibinfo {author} {\bibfnamefont
  {C.}~\bibnamefont {{Felser}}}, \bibinfo {author} {\bibfnamefont {B.~A.}\
  \bibnamefont {{Bernevig}}}, \ and\ \bibinfo {author} {\bibfnamefont
  {Z.}~\bibnamefont {{Wang}}},\ }\href@noop {} {\bibfield  {journal} {\bibinfo
  {journal} {arXiv}\ ,\ \bibinfo {eid} {arXiv:1909.04037}} (\bibinfo {year}
  {2019})}\BibitemShut {NoStop}%
\end{thebibliography}%


\begin{thebibliography}{7}%
\makeatletter
\providecommand \@ifxundefined [1]{%
 \@ifx{#1\undefined}
}%
\providecommand \@ifnum [1]{%
 \ifnum #1\expandafter \@firstoftwo
 \else \expandafter \@secondoftwo
 \fi
}%
\providecommand \@ifx [1]{%
 \ifx #1\expandafter \@firstoftwo
 \else \expandafter \@secondoftwo
 \fi
}%
\providecommand \natexlab [1]{#1}%
\providecommand \enquote  [1]{``#1''}%
\providecommand \bibnamefont  [1]{#1}%
\providecommand \bibfnamefont [1]{#1}%
\providecommand \citenamefont [1]{#1}%
\providecommand \href@noop [0]{\@secondoftwo}%
\providecommand \href [0]{\begingroup \@sanitize@url \@href}%
\providecommand \@href[1]{\@@startlink{#1}\@@href}%
\providecommand \@@href[1]{\endgroup#1\@@endlink}%
\providecommand \@sanitize@url [0]{\catcode `\\12\catcode `\$12\catcode
  `\&12\catcode `\#12\catcode `\^12\catcode `\_12\catcode `\%12\relax}%
\providecommand \@@startlink[1]{}%
\providecommand \@@endlink[0]{}%
\providecommand \url  [0]{\begingroup\@sanitize@url \@url }%
\providecommand \@url [1]{\endgroup\@href {#1}{\urlprefix }}%
\providecommand \urlprefix  [0]{URL }%
\providecommand \Eprint [0]{\href }%
\providecommand \doibase [0]{http://dx.doi.org/}%
\providecommand \selectlanguage [0]{\@gobble}%
\providecommand \bibinfo  [0]{\@secondoftwo}%
\providecommand \bibfield  [0]{\@secondoftwo}%
\providecommand \translation [1]{[#1]}%
\providecommand \BibitemOpen [0]{}%
\providecommand \bibitemStop [0]{}%
\providecommand \bibitemNoStop [0]{.\EOS\space}%
\providecommand \EOS [0]{\spacefactor3000\relax}%
\providecommand \BibitemShut  [1]{\csname bibitem#1\endcsname}%
\let\auto@bib@innerbib\@empty
\bibitem [{\citenamefont {Levin}\ \emph {et~al.}(2019)\citenamefont {Levin},
  \citenamefont {Grebenkemper}, \citenamefont {Pollock},\ and\ \citenamefont
  {Seshadri}}]{Levin2019}%
  \BibitemOpen
  \bibfield  {author} {\bibinfo {author} {\bibfnamefont {E.~E.}\ \bibnamefont
  {Levin}}, \bibinfo {author} {\bibfnamefont {J.~H.}\ \bibnamefont
  {Grebenkemper}}, \bibinfo {author} {\bibfnamefont {T.~M.}\ \bibnamefont
  {Pollock}}, \ and\ \bibinfo {author} {\bibfnamefont {R.}~\bibnamefont
  {Seshadri}},\ }\href {\doibase 10.1021/acs.chemmater.9b02594} {\bibfield
  {journal} {\bibinfo  {journal} {Chem. Mater.}\ }\textbf {\bibinfo {volume}
  {31}},\ \bibinfo {pages} {7151} (\bibinfo {year} {2019})}\BibitemShut
  {NoStop}%
\bibitem [{\citenamefont {Coelho}(2018)}]{Coelho2018}%
  \BibitemOpen
  \bibfield  {author} {\bibinfo {author} {\bibfnamefont {A.~A.}\ \bibnamefont
  {Coelho}},\ }\href {\doibase 10.1107/S1600576718000183} {\bibfield  {journal}
  {\bibinfo  {journal} {J. Appl. Crystallogr.}\ }\textbf {\bibinfo {volume}
  {51}},\ \bibinfo {pages} {210} (\bibinfo {year} {2018})}\BibitemShut
  {NoStop}%
\bibitem [{\citenamefont {Kaneko}, \citenamefont {Kanomata},\ and\
  \citenamefont {Shirakawa}(1987)}]{Kaneko1987a}%
  \BibitemOpen
  \bibfield  {author} {\bibinfo {author} {\bibfnamefont {T.}~\bibnamefont
  {Kaneko}}, \bibinfo {author} {\bibfnamefont {T.}~\bibnamefont {Kanomata}}, \
  and\ \bibinfo {author} {\bibfnamefont {K.}~\bibnamefont {Shirakawa}},\ }\href
  {\doibase 10.1143/JPSJ.56.4047} {\bibfield  {journal} {\bibinfo  {journal}
  {J. Phys. Soc. Jpn.}\ }\textbf {\bibinfo {volume} {56}},\ \bibinfo {pages}
  {4047} (\bibinfo {year} {1987})}\BibitemShut {NoStop}%
\bibitem [{\citenamefont {Fruchart}\ and\ \citenamefont
  {Bertaut}(1978)}]{Fruchart1978}%
  \BibitemOpen
  \bibfield  {author} {\bibinfo {author} {\bibfnamefont {D.}~\bibnamefont
  {Fruchart}}\ and\ \bibinfo {author} {\bibfnamefont {F.}~\bibnamefont
  {Bertaut}},\ }\href {\doibase 10.1143/JPSJ.44.781} {\bibfield  {journal}
  {\bibinfo  {journal} {J. Phys. Soc. Jpn.}\ }\textbf {\bibinfo {volume}
  {44}},\ \bibinfo {pages} {781} (\bibinfo {year} {1978})}\BibitemShut
  {NoStop}%
\bibitem [{\citenamefont {Heyd}, \citenamefont {Scuseria},\ and\ \citenamefont
  {Ernzerhof}(2006)}]{Heyd2003}%
  \BibitemOpen
  \bibfield  {author} {\bibinfo {author} {\bibfnamefont {J.}~\bibnamefont
  {Heyd}}, \bibinfo {author} {\bibfnamefont {G.~E.}\ \bibnamefont {Scuseria}},
  \ and\ \bibinfo {author} {\bibfnamefont {M.}~\bibnamefont {Ernzerhof}},\
  }\href {\doibase 10.1063/1.2204597} {\bibfield  {journal} {\bibinfo
  {journal} {J. Chem. Phys.}\ }\textbf {\bibinfo {volume} {124}},\ \bibinfo
  {pages} {219906} (\bibinfo {year} {2006})}\BibitemShut {NoStop}%
\bibitem [{\citenamefont {Dudarev}\ \emph {et~al.}(1998)\citenamefont
  {Dudarev}, \citenamefont {Botton}, \citenamefont {Savrasov}, \citenamefont
  {Humphreys},\ and\ \citenamefont {Sutton}}]{Dudarev1998}%
  \BibitemOpen
  \bibfield  {author} {\bibinfo {author} {\bibfnamefont {S.~L.}\ \bibnamefont
  {Dudarev}}, \bibinfo {author} {\bibfnamefont {G.~A.}\ \bibnamefont {Botton}},
  \bibinfo {author} {\bibfnamefont {S.~Y.}\ \bibnamefont {Savrasov}}, \bibinfo
  {author} {\bibfnamefont {C.~J.}\ \bibnamefont {Humphreys}}, \ and\ \bibinfo
  {author} {\bibfnamefont {A.~P.}\ \bibnamefont {Sutton}},\ }\href {\doibase
  10.1103/PhysRevB.57.1505} {\bibfield  {journal} {\bibinfo  {journal} {Phys.
  Rev. B}\ }\textbf {\bibinfo {volume} {57}},\ \bibinfo {pages} {1505}
  (\bibinfo {year} {1998})}\BibitemShut {NoStop}%
\bibitem [{\citenamefont {Antonov}\ \emph {et~al.}(2007)\citenamefont
  {Antonov}, \citenamefont {Harmon}, \citenamefont {Yaresko},\ and\
  \citenamefont {Shpak}}]{Antonov2007}%
  \BibitemOpen
  \bibfield  {author} {\bibinfo {author} {\bibfnamefont {V.~N.}\ \bibnamefont
  {Antonov}}, \bibinfo {author} {\bibfnamefont {B.~N.}\ \bibnamefont {Harmon}},
  \bibinfo {author} {\bibfnamefont {A.~N.}\ \bibnamefont {Yaresko}}, \ and\
  \bibinfo {author} {\bibfnamefont {A.~P.}\ \bibnamefont {Shpak}},\ }\href
  {\doibase 10.1103/PhysRevB.75.165114} {\bibfield  {journal} {\bibinfo
  {journal} {Phys. Rev. B}\ }\textbf {\bibinfo {volume} {75}},\ \bibinfo
  {pages} {165114} (\bibinfo {year} {2007})}\BibitemShut {NoStop}%
\end{thebibliography}%

\end{document}


\title[Supporting Material: Weyl nodes in Mn$_3$ZnC]{Supporting Material: Weyl nodes and magnetostructural instability in 
antiperovskite Mn$_3$ZnC}

\author{S. M. L. Teicher}
\affiliation{Materials Department and Materials Research Laboratory, University of California, Santa Barbara 93106, USA}
\email{steicher@ucsb.edu}
\author{I. K. Svenningsson}
\affiliation{Materials Research Laboratory, University of California, Santa Barbara 93106, USA}
\affiliation{Department of Applied Physics, Chalmers University of Technology 412 96, Sweden}
\author{L. M. Schoop}
\affiliation{Department of Chemistry, Princeton University, Princeton 08540, USA}
\author{R. Seshadri}
\affiliation{Materials Department and Materials Research Laboratory, University of California, Santa Barbara 93106, USA}

\date{\today}

\maketitle

\section{Additional Details: Experimental}

\begin{figure}[H]
\centering
\includegraphics[width=0.45\textwidth]{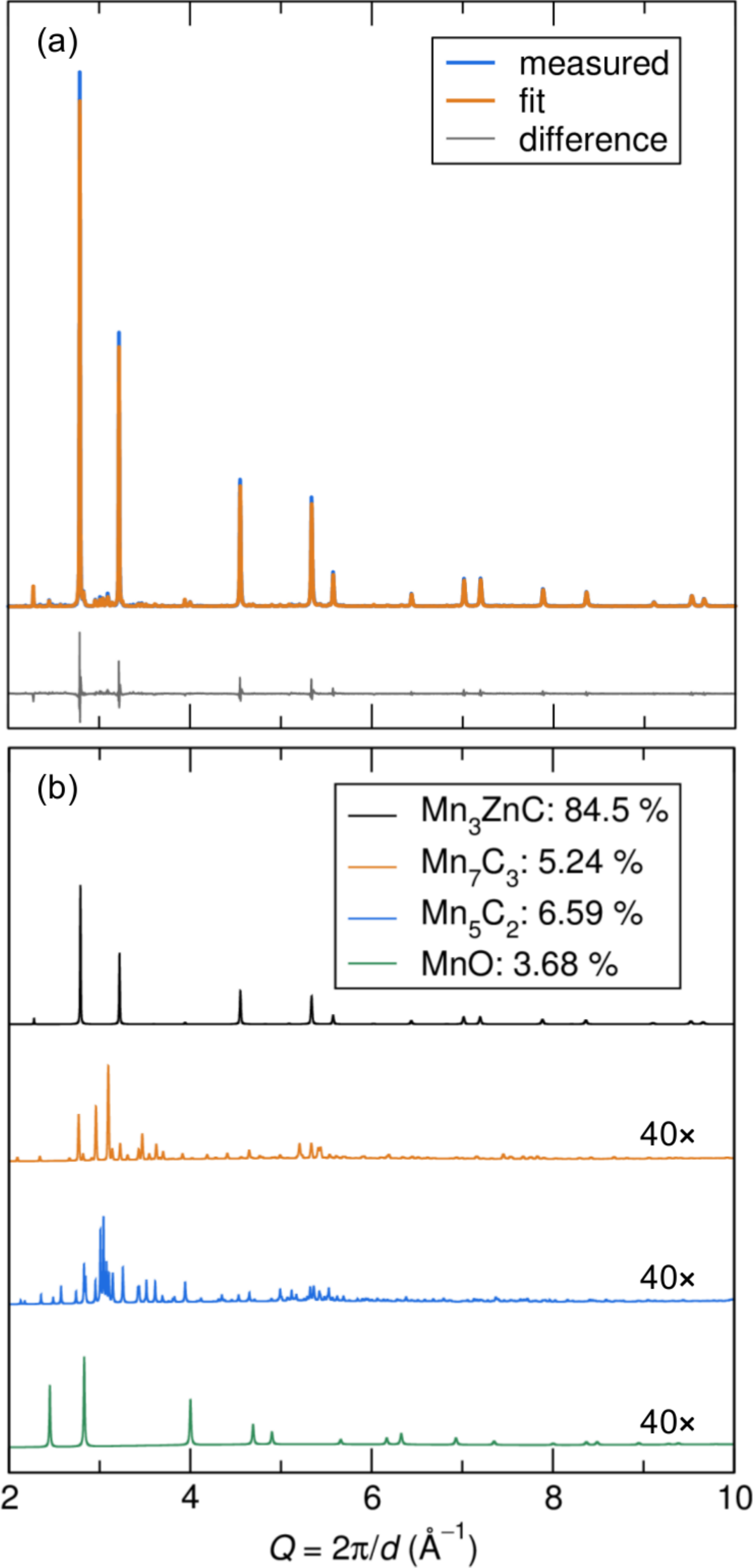}
\caption{\label{fig:FigureS1} Synchrotron X-ray diffraction data from a microwave sample. A Rietveld fit is shown in (a). (b) shows the majority Mn$_3$ZnC and minority manganese carbide and oxide contributions in our Rietveld model. Minority phase peak intensities are increased 40$\times$ in (b) for visual clarity.}
\end{figure}

We produced and measured the magnetic properties of three samples of Mn$_3$ZnC that appeared substantially pure in laboratory X-ray diffraction (PANalytical Empyrean powder diffractometer). In addition to the conventional synthesis with added zinc described in the text, we tried a straight stoichiometric conventional synthesis and an original microwave-assisted synthesis, also starting from stoichiometric precursors. Microwave-assisted synthesis often involves the use of activated charcoal as a susceptor since it is a strong microwave absorber; microwave heating was explored because it might change the reaction by locally heating carbon granules.\cite{Levin2019} We found that using activated charcoal (Sigma-Aldrich, DARCO, 12-20 mesh), rather than the graphite employed in the conventional synthesis, substantially improved microwave sample purity. The microwave-assisted synthesis involved sealing mixed Mn, Zn, and C powders in a carbonized, vacuum-sealed vitreous silica tube, placing this tube in a larger alumina crucible filled with activated charcoal, surrounding the tube / crucible with a protective enclosure of alumina insulation, and microwaving at 960\,W for 3 minutes in a commercial Panasonic microwave.\cite{Levin2019} We performed synchrotron X-ray diffraction at the 11-BM beamline at Argonne National Laboratory in order to verify the purity of the microwave samples since no prior studies on antiperovskite carbides have pursued this synthesis mechanism. These results are shown in Figure \ref{fig:FigureS1}. A Rietveld refinement performed in TOPAS\cite{Coelho2018} suggests that our microwave sample has small Mn$_7$C$_3$, Mn$_5$C$_2$, and MnO impurities, which may also be present in the conventionally-synthesized samples (though likely in smaller quantities). Both the conventionally-synthesized sample with no added zinc and the microwave sample had larger Mn:Zn ratios than the sample reported in the text as measured by wavelength-dispersive X-ray fluorescence, 3.8 and 4.6, respectively. We do not assign carbon concentration based on X-ray fluorescence due to weak signal and poor peak matching. The measured magnetic transitions in these samples were also found to be less sharp than those reported in the main body of the text. Although microwave synthesis appeared to produce samples with softer magnetic transitions than conventional synthesis methods for Mn$_3$ZnC---an indicator of poorer sample quality---this method is able to produce substantially pure powders in the antiperovskite structure within minutes and may be of interest for future studies on antiperovskite carbide family members.

\section{Additional Details: Computational}

\begin{table}[!b]
\begin{tabular}{c c c c}
    \toprule
    & expt. & non-SP & SP  \\
    \midrule
    $a$ (\AA) & $3.93$ & $3.78$ & $3.87$ \\
    $\mu_\textrm{Mn}$ ($\mu_B$) & $1.3$ & ($0.00$) & $2.41$ \\
    \\
    ICOHPs (eV) & & & \\
    Mn-Mn & & $-1.84$ & $-1.57$ \\
    Mn-Zn & & $-1.28$ & $-1.91$ \\
    Mn-C & & $-10.0$ & $-9.57$ \\
    $E_\textrm{DFT}$ (eV) & & & \\
     & & $0.714$ & $0.00$ \\
    \bottomrule
\end{tabular}
\caption{\label{tab:table1} Comparison of structure parameters and bonding in the experimental structure and the simulated non-spin polarized and spin polarized states for the ferromagnetic phase. Experimental lattice parameters and magnetic moments are sourced from \cite{Kaneko1987a} and \cite{Fruchart1978}, respectively. DFT energies are referenced to the ferromagnetic room temperature state.}
\end{table}

Tables \ref{tab:table1} and \ref{tab:table2} summarize the structure parameters and bonding in our simulations of the cubic ferromagnetic and noncollinear tetragonal structures, respectively. Compared to experiment, the lattice parameters are low and the magnetic moments high. This is consistent with the idea of this material as an itinerant magnet whose manganese moments are slightly over-localized by our simulation. We do not discuss modeling the system with hybrid functionals such as HSE06 or Hubbard $U$ corrections commonly used for more localized magnetic systems as, both in general and in this system specifically, these techniques readily result in Mn moments in excess of 3$\mu_B$ which are inconsistent with experiment.\cite{Heyd2003,Dudarev1998} The integrated COHPs, which can be viewed as a proxy for bond strength, confirm the energetic stabilization mechanism for the paramagnetic to ferromagnetic transition described in the text: spin polarization in the cubic structure strengthens Mn-Zn bonding at the expense of Mn-C and Mn-Mn bonding. The ferromagnetic configuration is favored over the nonmagnetic configuration by an energy of 0.714 eV per unit cell.

\begin{table}[!b]
\begin{tabular}{c c c}
    \toprule
    & expt. & NCL \\
    \midrule
    $a$ (\AA) full cell & $5.54$ & $5.49$ \\
    $c$ (\AA) full cell & $7.79$ & $7.66$ \\
    $a$ (\AA) FM cell & $3.92$ & $3.89$ \\
    $c$ (\AA) FM cell & $3.90$ & $3.83$ \\
    $\mu_\textrm{Mn$_1$}$ ($\mu_B$) & $1.6$ & $2.09$ \\
    $\mu_\textrm{Mn$_2$}$ ($\mu_B$) & $2.7$ & $2.47$ \\
    \\
    $E_\textrm{DFT}$ (eV) & & \\
    & & $0.03$ \\
    \bottomrule
\end{tabular}
\caption{\label{tab:table2} Comparison of structure parameters and bonding in the experimental structure and the non-collinear simulation for the noncollinear ferrimagnetic ground state. Experimental lattice parameters and magnetic moments are sourced from \cite{Antonov2007} and \cite{Fruchart1978}, respectively. The DFT energy is referenced to the ferromagnetic room temperature state.}
\end{table}

Despite ample experimental evidence that the ground state of this system is the noncollinear structure, the calculated DFT energy of this state is slightly less favorable than the cubic ferromagnet. The energetic difference of 0.03\,eV per primitive cell is small with respect to the expected error in our simulation. A previous first-principles study within the LMTO-LSDA framework reported a similar discrepancy in which the ground state of Mn$_3$ZnC was found to be slightly less energetically favorable.\cite{Antonov2007} The lattice parameters of the low temperature structure are provided both for the full low-temperature cell and also referenced to the FM structure containing a single FM cell with one Mn-C octahedron. Comparing the latter values, the tetragonal structural distortion associated with the the ferromagnetic to ferrimagnetic transition can be seen to be relatively small in both theory and prior experiment. This small structural distortion is consistent with the modest measured $\Delta S_M$ values.

Additional figures address the total density of states of the cubic phase, the effect of spin orbit coupling on the bulk electronic structure, nodal lines, and topological surface states, the tight-binding model fitting, and the band structure of the low temperature structure (without band unfolding).

\begin{figure}[H]
\centering
\includegraphics[width=0.45\textwidth]{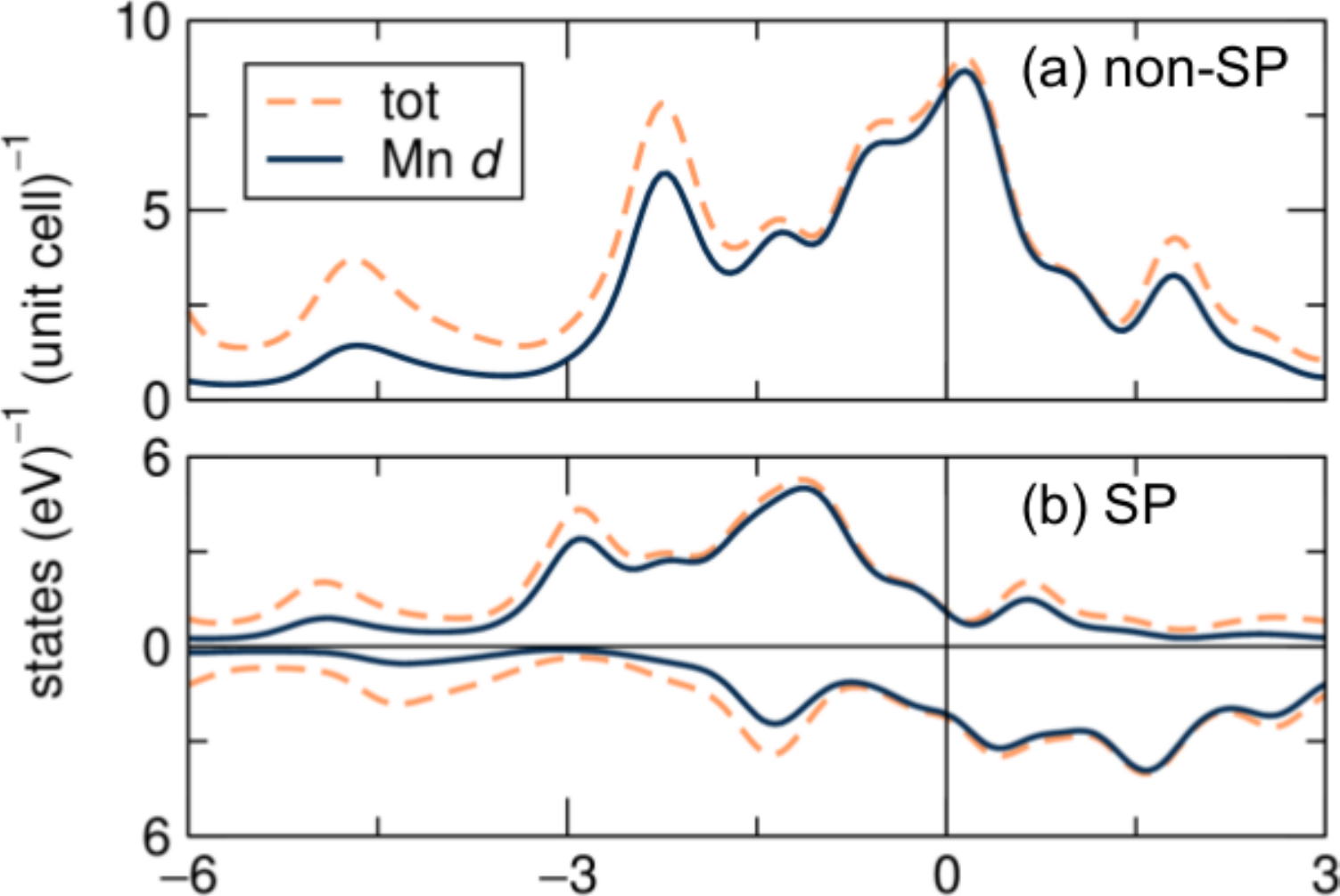}
\caption{\label{fig:FigureS2} Density of states plots comparing the total densities of states and projected Mn $d$ orbital contribution for (a) the non-spin polarized and (b) spin polarized simulations of the cubic structure. The main contributions in the region near the Fermi level arise from Mn $d$ states and the states at the Fermi level are almost exclusively Mn $d$.}
\end{figure}

\begin{figure}[H]
\centering
\includegraphics[width=0.45\textwidth]{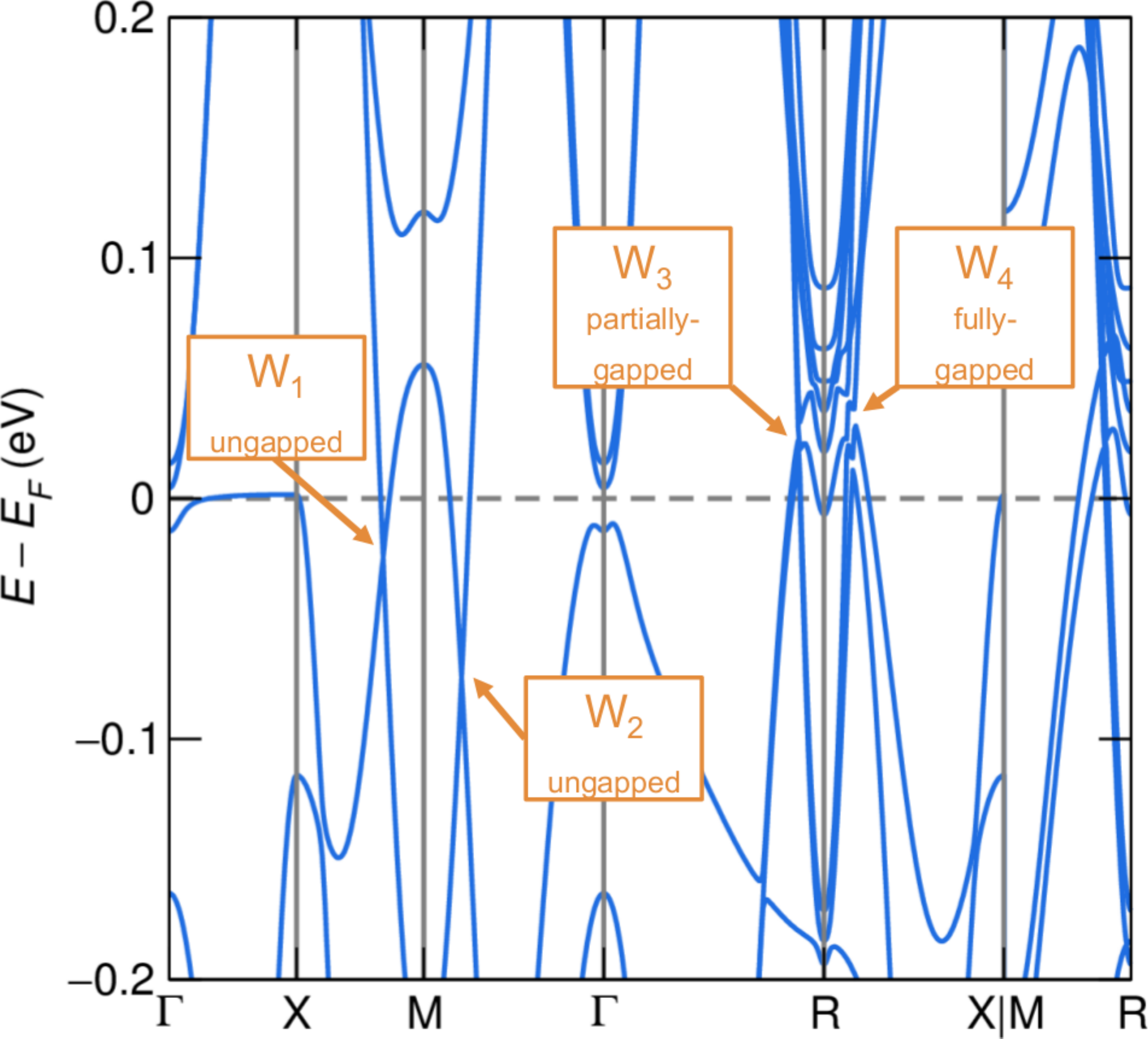}
\caption{\label{fig:FigureS3} Bulk band structure with spin-orbit coupling. As described in the text, the $W_1$ and $W_2$ Weyl crossings on the $k_z=0$ plane are preserved while $W_4$ and all but one of the $W_3$ band crossings gap out.}
\end{figure}

\begin{figure}[H]
\centering
\includegraphics[width=0.9\textwidth]{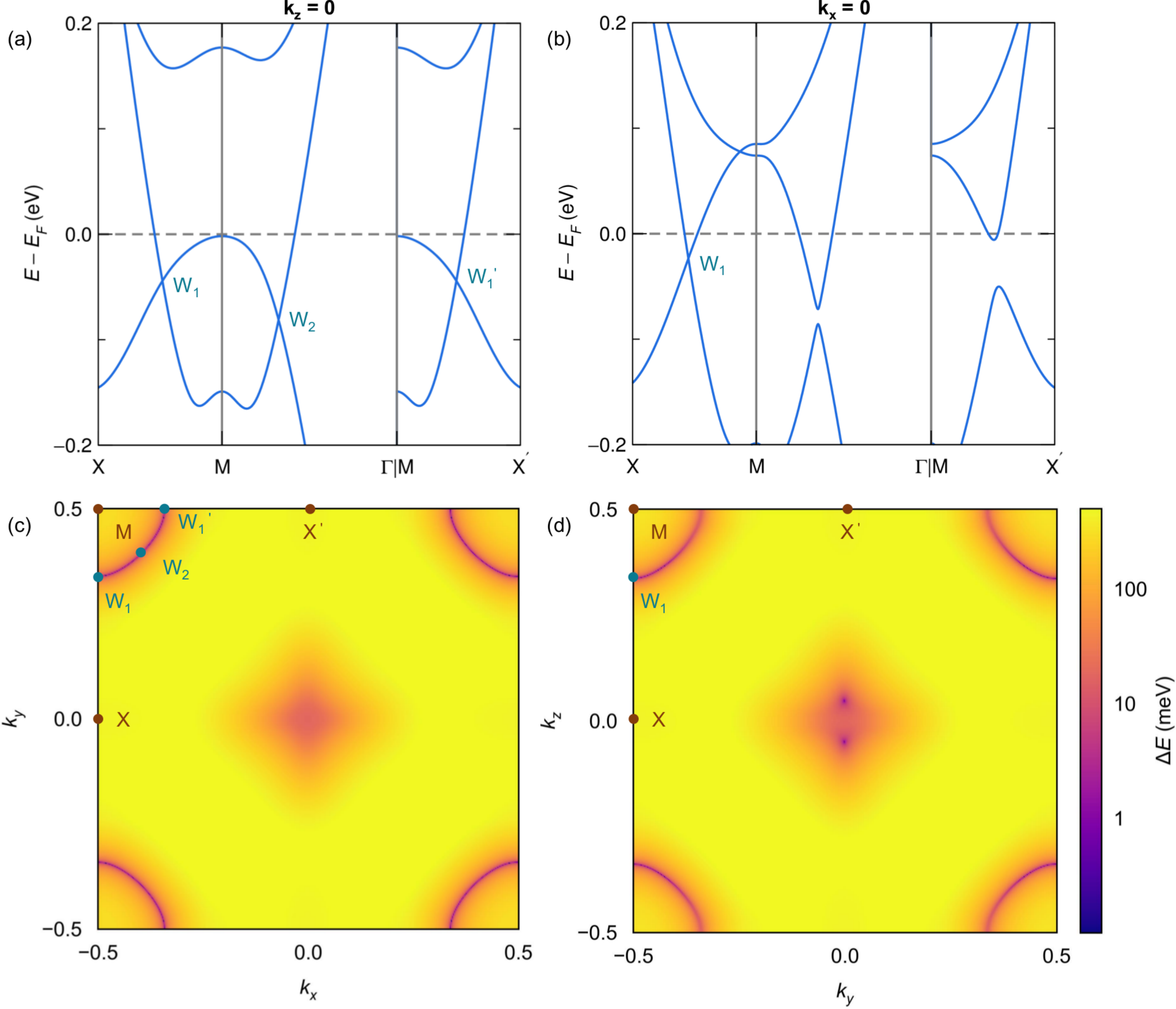}
\caption{\label{fig:FigureS4} Characterization of spin-orbit gapping of the nodal lines about $M$. In the $k_z=0$ plane, the [001] magnetic ordering does not break mirror symmetry, and the nodal line is preserved. In the $k_x=0$ plane, the magnetic ordering breaks symmetry, resulting in a distinction between $X$ and $X^\prime$ points and a subtle gapping of the nodal line at all points except the $W_1$ crossing along $M$-$X$. (a,b) band structures calculated on the $k_z=0$ and $k_x=0$ planes show protected and gapped nodes, respectively. The spin-orbit coupling strength was artificially increased by a factor of 3 in these simulations in order to accentuate the gaps for ease of viewing. (c,d) Color maps of the energy separation, $\Delta E$, between the two simulation-defined bands forming the $W_1,W_1^\prime,W_2$ nodes, calculated on a fine $k$-mesh using our tight binding models, show the protected and gapped nodal lines about $M$ on the $k_z=0$ and $k_x=0$ planes, respectively. The color is darker for all points at which the two bands come close in energy, including the Weyl nodal lines and an unrelated region immediately about $\Gamma$ at the center of the plots. You may notice the two sharp points near $\Gamma$ in (d), these are indeed additional Weyl points, $W_5$, detailed in Figure \ref{fig:FigureS5}.}
\end{figure}

\begin{figure}[H]
\centering
\includegraphics[width=0.45\textwidth]{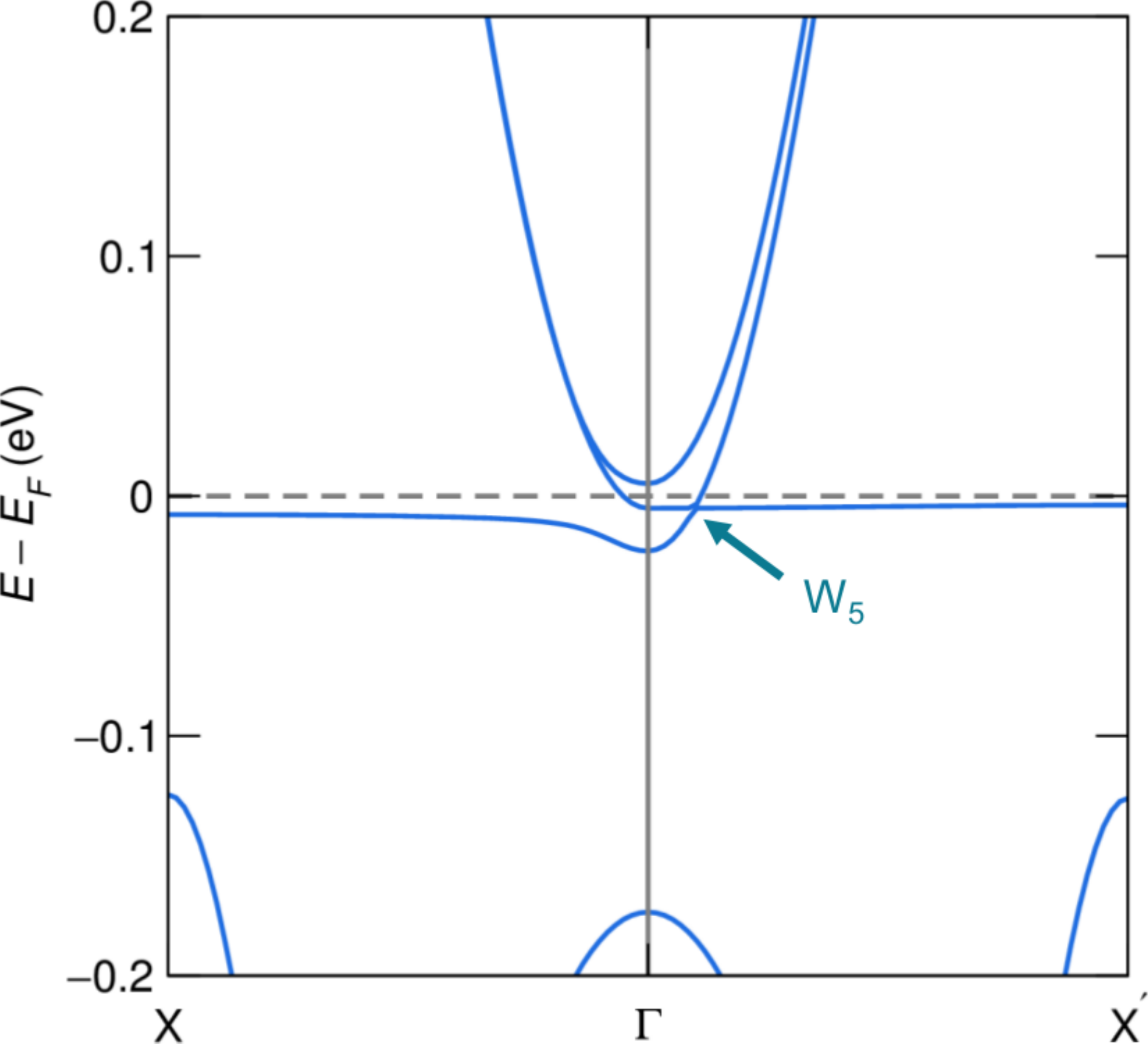}
\caption{\label{fig:FigureS5} Band structure near the $W_5$ nodes that result from the spin-orbit interaction, which splits the upper parabolic band at $\Gamma$, pushing it through the $\Gamma$-$X$ flat band. While symmetry-breaking associated with the [001] magnetic moment gaps this crossing in most directions, the crossing along $\Gamma$-$X^\prime$, parallel to the magnetization, is preserved. For the sake of succinctness, these Weyl nodes are neglected in the text. The energy scale associated with the $W_5$ nodes and their dispersion, about 0.01\,eV, is small with respect to that of the other Weyl nodes and is likely less relevant to the magnetostructural coupling than the bonding-driven Weyl crossings described in the text.}
\end{figure}

\begin{figure}[H]
\centering
\includegraphics[width=0.9\textwidth]{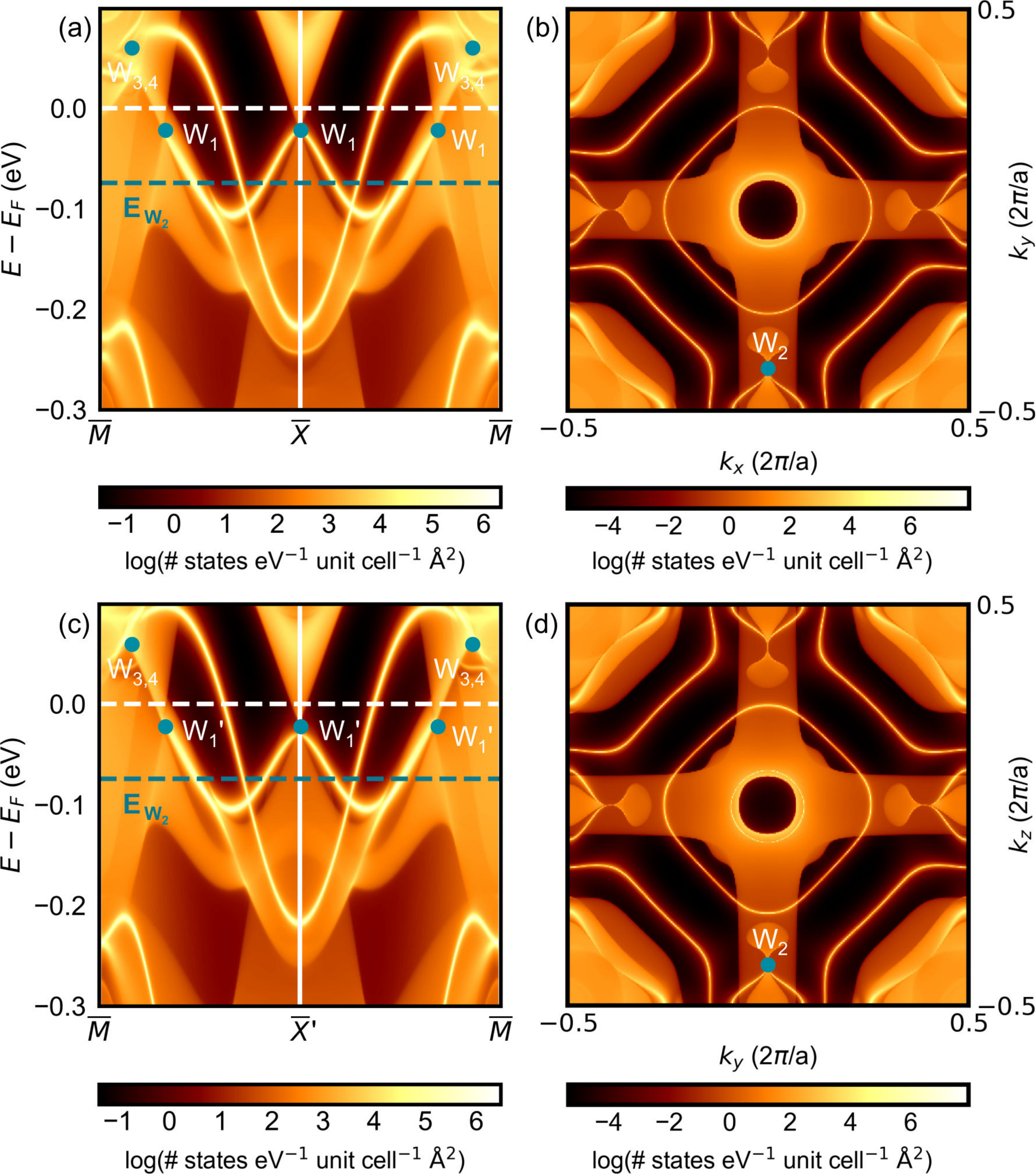}
\caption{\label{fig:FigureS6} Effect of spin-orbit coupling (SOC) on the Weyl and drumhead surface states. (a,b) show the dispersion of the surface density of states and Fermi arc calculations at the energy level of the $W_2$ node from the main text repeated for the spin-orbit simulation with projection on a (001) surface. Because the $k_z=0$ nodal line is symmetry protected, the surface states are very similar to those shown for the calculation without SOC. (c,d) the same calculations again for (100) surface states. Here, the symmetry breaking subtly gaps the Weyl nodes as can be seen in (c) (the surface density of states are shown on the $\overline{M}$-$\overline{X}^\prime$-$\overline{M}$ with maximal Weyl node gapping, rather than the$\overline{M}$-$\overline{X}$-$\overline{M}$ line). Magnetic ordering also introduces a slight anisotropy in the Weyl surface states as shown in (d). Overall, the weak SOC has a minor effect on the surface states.}
\end{figure}

\begin{figure}[H]
\centering
\includegraphics[width=0.45\textwidth]{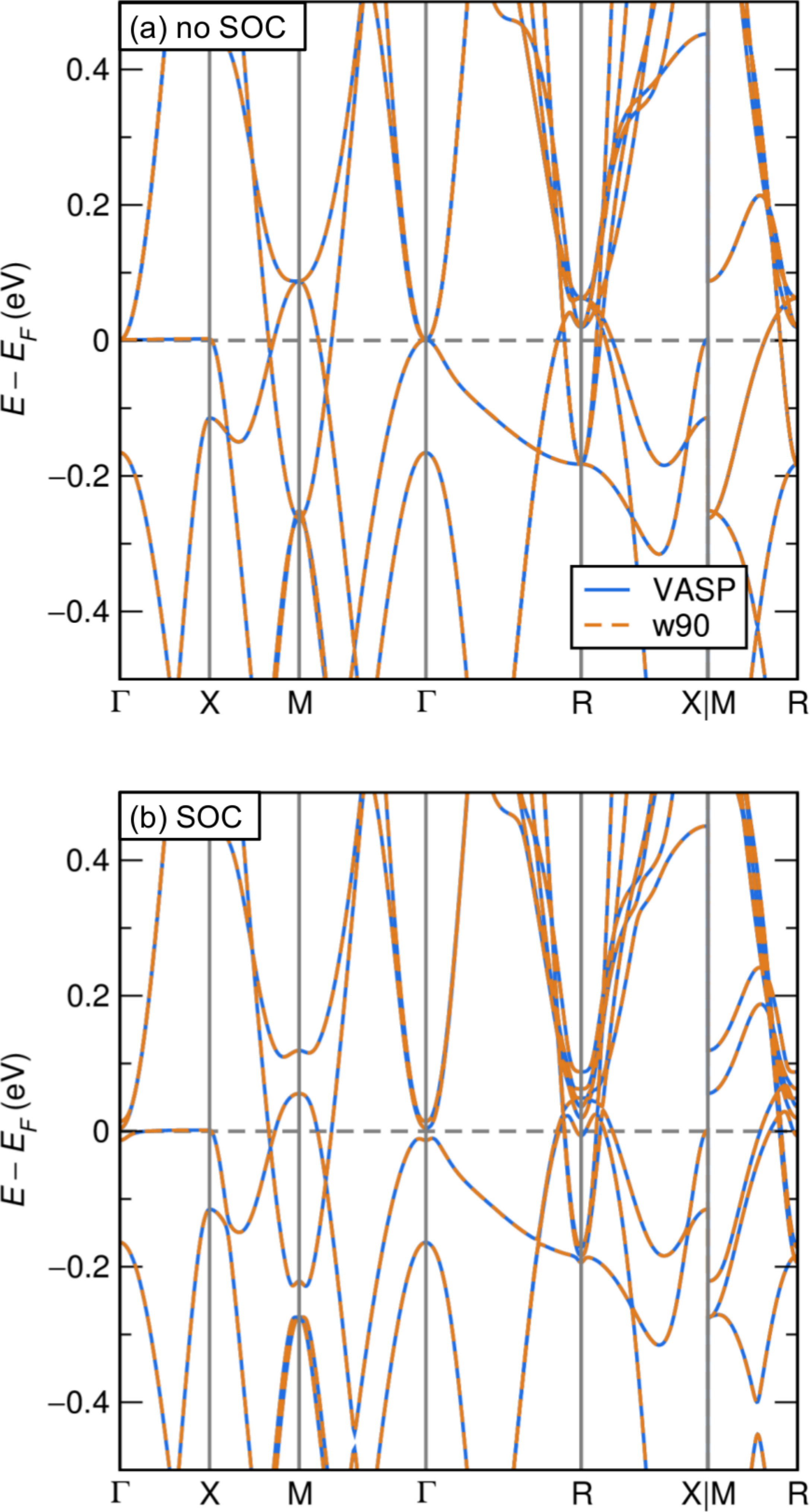}
\caption{\label{fig:FigureS7} Comparison between band structures calculated in VASP and using our Wannier-projected tight binding models for (a) the simulation without spin-orbit coupling and (b) the simulation with spin orbit-coupling. There is good agreement in the region of interest near the Fermi level.}
\end{figure}

\begin{figure}[H]
\centering 
\includegraphics[width=0.45\textwidth]{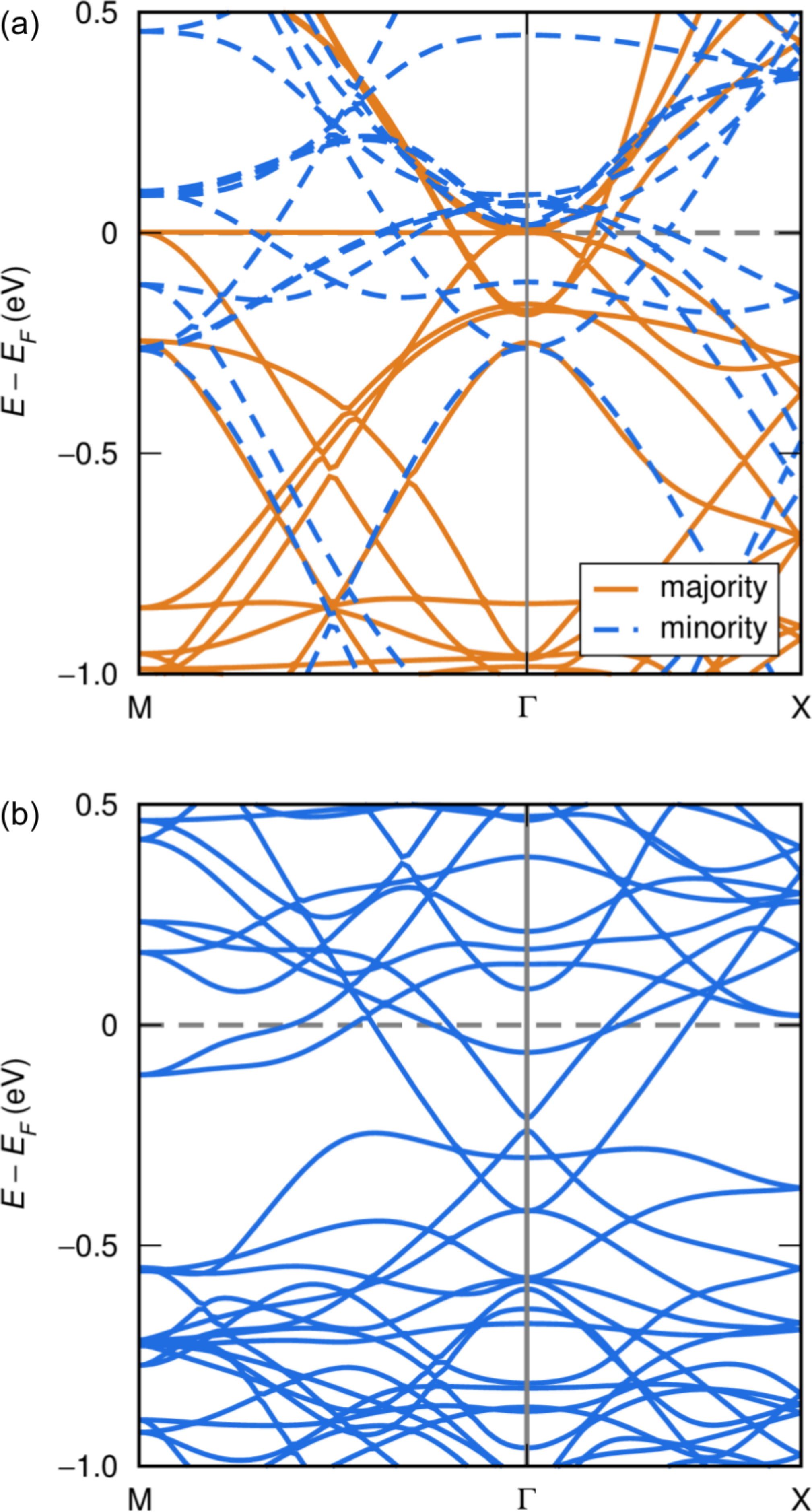}
\caption{\label{fig:FigureS8} Comparison of band structures calculated in the low temperature, tetragonal cell. (a) shows the spin polarized band structure calculated by expanding the ferromagnetic cubic cell into a supercell. The flat bands and Weyl crossings of the primitive cell are present as well as additional band crossings due to the folding of the Brillouin zone for this expanded cell. The flat bands are along $M$-$\Gamma$, rather than $X$-$\Gamma$, because of the 45$^\circ$ rotation of the cell axes relative to the primitive cell for the tetragonal structure described in the text. (b) shows the band structure of the noncollinear tetragonal cell without band unfolding, showing no flat bands and few Weyl nodes near the Fermi level.}
\end{figure}

\bibliography{my_bib}